%% file: main.tex
\documentclass[aps,prd,reprint,superscriptaddress,nofootinbib,longbibliography]{revtex4-2}

\usepackage{amsmath,amssymb}
\usepackage{graphicx}
\usepackage{booktabs}
\usepackage{siunitx}
\usepackage{orcidlink}
\usepackage{xcolor}
\usepackage{hyperref}
\usepackage{placeins}
\usepackage{lineno}

\hypersetup{colorlinks=true,linkcolor=blue,citecolor=blue,urlcolor=blue}

\newcommand{\numu}{\ensuremath{\nu_\mu}}
\newcommand{\nue}{\ensuremath{\nu_e}}
\newcommand{\nutau}{\ensuremath{\nu_\tau}}
\newcommand{\pizero}{\ensuremath{\pi^0}}
\newcommand{\Enu}{\ensuremath{E_\nu}}
\newcommand{\Ereco}{\ensuremath{E_\nu^{\mathrm{reco}}}}
\newcommand{\dmsq}{\ensuremath{\Delta m^2_{41}}}
\newcommand{\mfour}{\ensuremath{m_4}}
\newcommand{\Uefour}{\ensuremath{|U_{e4}|^2}}
\newcommand{\Umufour}{\ensuremath{|U_{\mu 4}|^2}}
\newcommand{\Uafour}{\ensuremath{|U_{\alpha 4}|^2}}
\newcommand{\sstt}{\ensuremath{\sin^2 2\theta}}
\newcommand{\srel}{\ensuremath{\sigma_{\mathrm{rel}}}}
\makeatletter
\newcommand{\rnd}[2]{{\def\ensuremath##1{##1}%
  \xdef\rnd@tmp{\noexpand\num[round-mode=places,round-precision=#1]{#2}}}\rnd@tmp}
\makeatother

\input{numbers.tex}

\graphicspath{{figures/}{figures/1_separation_nue_pi0/}{figures/2_energy_resolution/}{figures/3_sensitivity/}{figures/0_detector/}}

\begin{document}

\title{Sensitivity of the SHiP experiment at CERN to sterile-neutrino-induced oscillations}

\author{SHiP Collaboration}
\thanks{The complete list of authors and affiliations appears at the end of this article.}

\date{\today}

\begin{abstract}
The SHiP experiment, approved for the high-intensity beam-dump facility in the
ECN3 hall at the CERN SPS in the North Area, will record neutrino interactions both in its scattering
and neutrino detector (SND) and, about \SI{70}{m} further downstream, in its calorimeter system. We
show that the baseline SHiP detector configuration therefore forms a two-baseline experiment
for oscillations induced by a sterile neutrino. From a \textsc{Geant4}
simulation of \textsc{Genie}-generated interactions we derive the energy response of
both stations to $\nue$ and $\numu$ charged-current (CC) interactions and the
separation of $\nue$~CC events from the neutral-current $\pizero$ background,
and fold them into a near/far disappearance analysis with a $20\%$ flux-shape
uncertainty and a $1\%$ far/near normalisation uncertainty. For
$6\times10^{20}$ protons on target, the projected $95\%$~CL median sensitivities
are $\Umufour\simeq\UmuLimit$ at $\mfour\simeq\UmuLimitMass$~eV and
$\Uefour\simeq\UeLimit$ at $\mfour\simeq\UeLimitMass$~eV. In the muon channel
this lies below existing $\numu$-disappearance limits for
$\DnineLow\lesssim\mfour\lesssim\DnineHigh$~eV.
\end{abstract}

\maketitle

\section{Introduction}
\label{sec:intro}

\begin{figure*}[t]\centering
  \includegraphics[width=\textwidth]{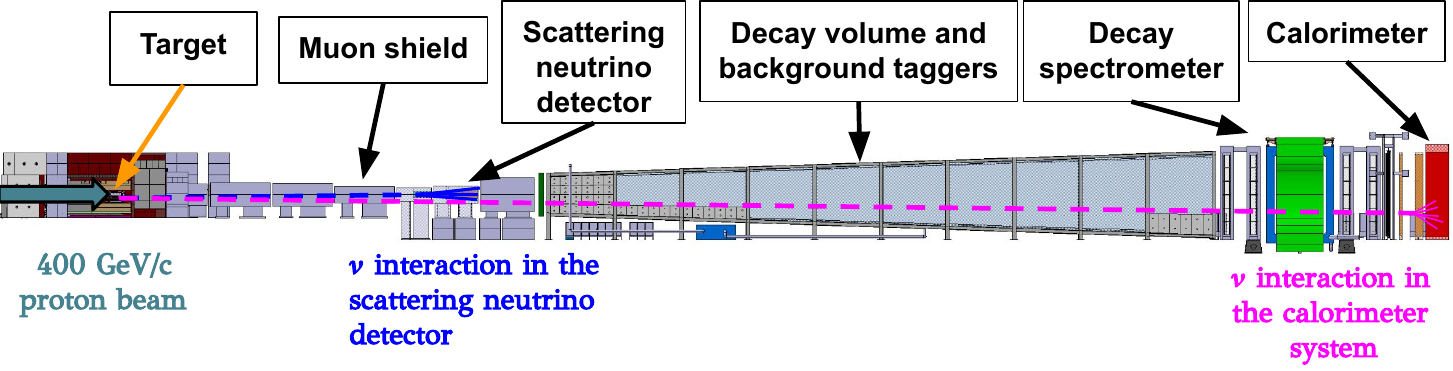}
  \caption{Layout of the SHiP experiment, with the neutrino interactions in the
  scattering and neutrino detector (SND) and in the calorimeter system, the near and far stations of
  this study.
  \label{fig:layout}}
\end{figure*}

SHiP is a general-purpose beam-dump experiment at the CERN SPS, designed to
search for feebly interacting particles and to study the interactions of the
intense flux of neutrinos of all flavours produced by \SI{400}{GeV} protons in
a dense tungsten target~\cite{SHiP_ECN3,Climescu:2026omx}. Downstream of the
target and the muon shield, a scattering and neutrino detector (SND) instrumented for
vertexing, tracking and energy measurement records neutrino interactions for
the neutrino-physics programme. About \SI{70}{m} further downstream, after the
decay volume and the spectrometer of the hidden-sector search, a calorimeter
system provides particle identification for that search, while also recording neutrino
interactions (Fig.~\ref{fig:layout}).

Light sterile neutrinos have been invoked to explain several anomalous results:
the excesses of electron-like events in LSND and
MiniBooNE~\cite{LSND,MiniBooNE2021}, the deficit of reactor
antineutrinos~\cite{ReactorAnomaly}, and the gallium
anomaly~\cite{BEST2022}. These anomalies have been the subject of extensive investigation~\cite{Dentler:2018sju,SterileWhitePaper}. Searches for a fourth, mostly sterile mass eigenstate also probe the thermal history of the Universe before Big Bang nucleosynthesis. In particular, decays of light particles into charged leptons and neutrinos can lead to a late reheating, as late as a few hundred keV, while remaining consistent with the cosmic microwave background and the light-element abundances and with a limited impact on structure formation~\cite{Akita:2026sqe}.
This impact is now being tested by large-scale-structure and CMB surveys: within $\Lambda$CDM and in combination with CMB data from Planck~\cite{Planck2018} and ACT~\cite{ACT:DR6lensing}, DESI constrains the summed neutrino mass to $\sum m_\nu<0.064$~eV and the effective number of relativistic species to $N_\mathrm{eff}=3.23^{+0.35}_{-0.34}$ at $95\%$~CL~\cite{DESI:DR2BAO,DESI:DR2nu}, constraints that depend on the assumed cosmological model, while Euclid~\cite{Euclid:Overview,Euclid:Neutrinos}, the Rubin Observatory LSST~\cite{LSST:Ivezic2019,LSST:NuMass} and the Simons Observatory~\cite{SO:2019,SO:2025} will measure the growth of structure at the percent level and further constrain additional light relics.
In the standard cosmological model, a sterile mass eigenstate with
$\dmsq\gtrsim\SI{10}{eV^2}$ and mixing at the $10^{-3}$--$10^{-2}$ level would be in
strong tension with observations~\cite{Planck2018}. In
this mass range the mixing with $\numu$ is constrained by short-baseline
accelerator experiments~\cite{CDHSW,CCFR,MINOS_sterile,MBSB2012,NOvA2026,ICARUS:2026}
and by atmospheric neutrinos~\cite{SuperK2015,IceCubeDC2024}, and
the mixing with $\nue$ by tritium $\beta$
decay~\cite{Mainz2013,Troitsk2014,KATRIN:2025lph}. It should be noted that the oscillation signature
studied here is distinct from the searches for the decay of heavy neutral
leptons in the SHiP decay volume, which address masses in the GeV range.

A sterile state with $\dmsq\sim10^3$~\si{eV^2} oscillates over a few tens of
metres at the energies of the SHiP neutrino beam (Sec.~\ref{sec:model}), the
scale of the distances between the SHiP target, the SND and the calorimeter
system. Comparing the two stations therefore measures the oscillation at two
baselines. Here we show that this measurement can be performed with competitive sensitivity in the baseline SHiP
detector configuration, with the SND as the near station and the calorimeter system as the far station, and we
derive the response of both stations from simulation.

Two earlier studies addressed sterile-neutrino oscillations at SHiP.
Reference~\cite{Choi2024} studied $\nutau$ disappearance, which constrains
$|U_{\tau4}|^2$, with the emulsion-based neutrino detector of an earlier SHiP
design, in which the $\tau$ lepton is identified from its decay topology. In the
simulated silicon--tungsten target most $\tau$ leptons decay before reaching
the next silicon plane, so that their decay topology cannot be reconstructed event by event,
and the calorimeter system was not designed for this purpose. The $\nutau$
channel is therefore not studied here and is left to future work. Reference~\cite{Choi2025} extended the study
to all three flavours, in appearance and disappearance, with a near detector at
\SI{27}{m} and an additional far detector at \SI{120}{m} of the same acceptance,
generic detector responses and normalisation uncertainties. The present work
uses the baseline SHiP detector configuration, detector responses derived from simulation and a
flux-shape uncertainty in addition to the normalisation uncertainties; the
sensitivities of the two studies are compared in Sec.~\ref{sec:sens}.

\section{Sterile neutrinos in the 3+1 model}
\label{sec:model}

In the 3+1 model the three active flavour states mix with a fourth, sterile
state through a unitary $4\times4$ matrix $U$,
$\nu_\alpha=\sum_{i=1}^{4}U_{\alpha i}\,\nu_i$ with $\alpha=e,\mu,\tau,s$. The
fourth mass eigenstate $\nu_4$ is mostly sterile, $|U_{s4}|^2\simeq1$, and
admixed to the active flavours through the small elements $|U_{\alpha4}|^2$.
For $\dmsq\gg|\Delta m^2_{31}|$ the standard three-flavour oscillations have no
time to develop over the baselines of SHiP ($|\Delta m^2_{31}|L/E\sim10^{-6}$
for $L\sim\SI{100}{m}$ and $E\sim\SI{40}{GeV}$), and the oscillation
probabilities in vacuum depend only on the fourth column of
$U$~\cite{PDG,SterileWhitePaper},
\begin{align}
P(\nu_\alpha\!\to\!\nu_\alpha)
&= 1-\sstt_{\alpha\alpha}\sin^2\Delta_{41},
\label{eq:posc}\\
P(\nu_\alpha\!\to\!\nu_\beta)
&= 4\,|U_{\alpha4}|^2|U_{\beta4}|^2\sin^2\Delta_{41},
\label{eq:papp}
\end{align}
for $\beta\neq\alpha$, with $\Delta_{41}=1.267\,\dmsq L/E$ for the baseline $L$
in \si{km}, $E$ in \si{GeV} and \dmsq{} in \si{eV^2}, and the disappearance
amplitude
\begin{equation}
\sstt_{\alpha\alpha}=4\,\Uafour\left(1-\Uafour\right)
=4\,\Uafour\sum_{\beta\neq\alpha}|U_{\beta4}|^2 ,
\label{eq:s2t}
\end{equation}
where the second equality follows from the unitarity of $U$. Matter effects are negligible: the
potential of the traversed material is many orders of magnitude below
$\dmsq/2E$. As in the external limits we compare with, each channel is analysed
with a single non-zero element, $\Umufour$ for $\numu$ and $\Uefour$ for
$\nue$. With both non-zero, $\numu\to\nue$ appearance, Eq.~\eqref{eq:papp},
would refill about $1\%$ of the $\nue$ deficit~\cite{Choi2025}, which is
neglected; the CPT-conserving model gives the same probabilities for neutrinos
and antineutrinos, which are therefore not separated.

The sensitivity of an experiment is set by its ratio $L/E$. The first
oscillation maximum lies at \mbox{$\dmsq\simeq1.24\,\si{eV^2}/(L/E\,[\si{km/GeV}])$}:
for a mean energy of about \SI{40}{GeV}, at about $5\times10^2$~\si{eV^2} for
the calorimeter system and $2\times10^3$~\si{eV^2} for the SND. Far below the
maximum the oscillation has no effect; around it, the deficit depends on energy
and differs between the two stations, which is where a two-baseline comparison
is most sensitive. Far above it, the oscillation is averaged over the energy
resolution and the interaction depth, leaving a constant deficit
$\tfrac12\sstt_{\alpha\alpha}$ that is degenerate with the flux normalisation.
Experiments at $L/E\sim0.1$--\SI{1}{km/GeV}, such as the Short-Baseline Neutrino
programme at Fermilab~\cite{SBN_prospects,ICARUS:2026}, are in this averaged
regime at both of their detectors above $\dmsq\sim10^2$~\si{eV^2}, where the
SHiP baselines still resolve the oscillation in energy.

Since the dominant active neutrino eigenstates are much lighter than $\nu_4$, its mass is
$\mfour\simeq\sqrt{\dmsq}$, and the same exclusion can be expressed in the
mass--mixing plane $(\mfour,\Uafour)$, the plane used for heavy neutral leptons. At
the masses considered here the fourth state is far too light and too long-lived
to decay in the apparatus, and the constraint comes only from the disappearance
of active flavours.

\section{Simulation, detectors and event samples}
\label{sec:sim}

Neutrino interactions of all flavours are generated with \textsc{Genie}~3.6.2, tune
G21\_11a\_00\_000~\cite{GENIE}, for $4\times10^{19}$ protons on target (PoT),
the nominal annual SHiP exposure, and scaled to the $6\times10^{20}$~PoT of the full
15-year programme. The final states are transported through a standalone
\textsc{Geant4} simulation~\cite{Geant4}, injected at the true vertex.

\subsection{Scattering and neutrino detector}
\label{sec:sdgeom}

\begin{figure}[t]\centering
  \scalebox{-1}[1]{\includegraphics[width=\columnwidth]{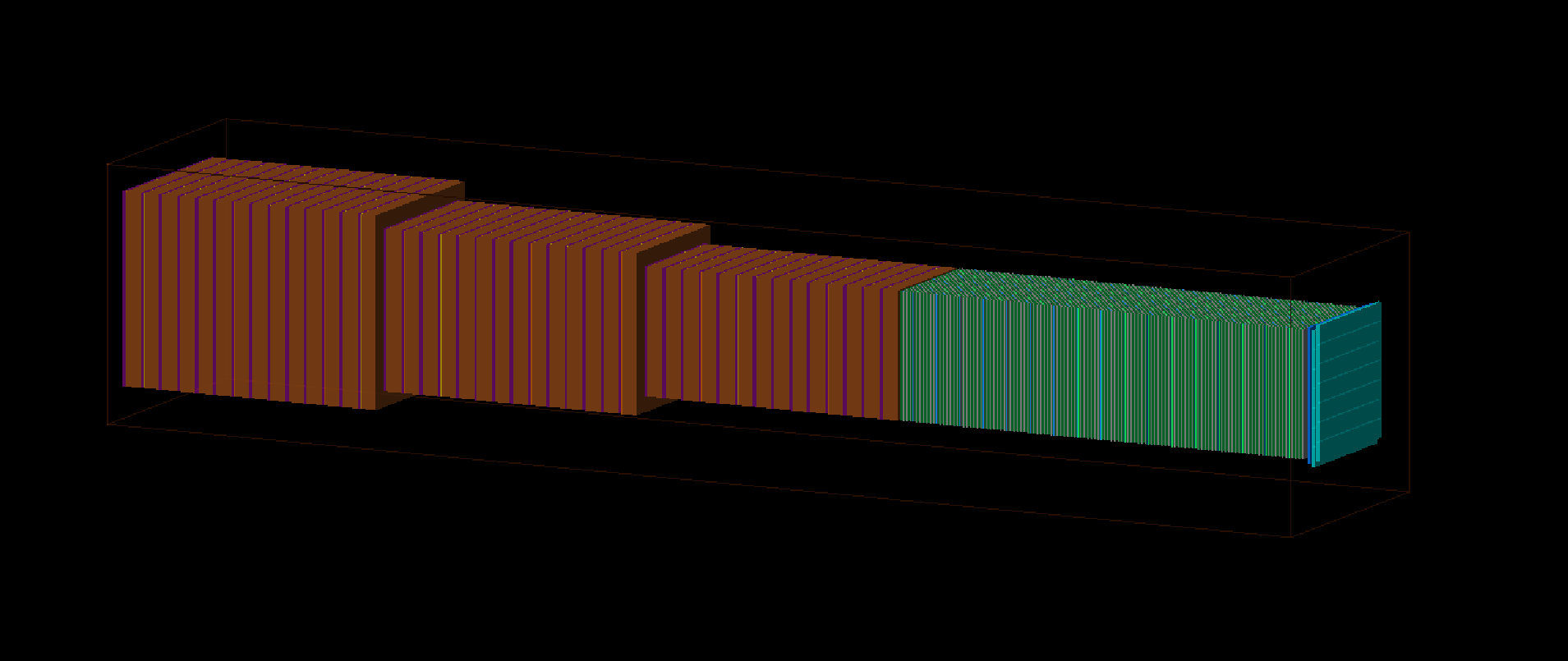}}
  \caption{Simulated geometry of the scattering and neutrino detector: the
  silicon--tungsten target [green] and the magnetised iron [orange] sections. The beam enters from the left.
  \label{fig:sd_geom}}
\end{figure}

The simulated SND (Fig.~\ref{fig:sd_geom}) is a version of the design in which
the detector sits inside the last magnet of the muon shield, whose field of
\SI{1.9}{T} magnetises its iron. Downstream of a plastic-scintillator veto
($\rnd{2}{\SNDVetoZmin}<z<\rnd{2}{\SNDVetoZmax}$~m), it consists of a $40\times\SI{40}{cm^2}$
silicon--tungsten target ($\rnd{2}{\SNDTargetZmin}<z<\rnd{2}{\SNDTargetZmax}$~m) of
$\SNDTargetPlates$ tungsten plates of $\SNDPlateThick$~mm, each followed by an
$x$ and a $y$ plane of silicon strips of \SI{75}{\micro m} pitch, and of three
magnetised iron sections of 40, 50 and \SI{60}{cm} transverse size
($\rnd{2}{\SNDFeZminOne}$--$\rnd{2}{\SNDFeZmaxOne}$, $\rnd{2}{\SNDFeZminTwo}$--$\rnd{2}{\SNDFeZmaxTwo}$ and
$\rnd{2}{\SNDFeZminThree}$--$\rnd{2}{\SNDFeZmaxThree}$~m). Each section holds
$\SDPlanesPerSection$ iron plates of $\SDIronSampling$~cm, interleaved with
sensitive planes that combine, for tracking, two $\SNDFibreLayerThick$~mm layers
of scintillating fibres of $\SNDFibreDiam$~\si{\micro m} diameter and, for
calorimetry, a $\SNDTileThick$~mm layer of
$\SNDTileSize\times\SNDTileSize$~\si{cm^2} scintillator tiles. The fine segmentation of the target separates $\nue$
charged-current (CC) interactions from neutral-current (NC) interactions
producing a $\pizero$ (NC-$\pizero$ events, Sec.~\ref{sec:nuesep}), and the magnetised iron
measures the muon momentum (Sec.~\ref{sec:numunear}). The detector layout is
still evolving, which may lead to some variation in the exact sensitivity
achieved; however, the underlying two-baseline principle and the qualitative
conclusions of this study are not expected to be significantly affected.

\subsection{Calorimeter system}
\label{sec:calogeom}

\begin{figure}[t]\centering
  \scalebox{-1}[1]{\includegraphics[width=\columnwidth]{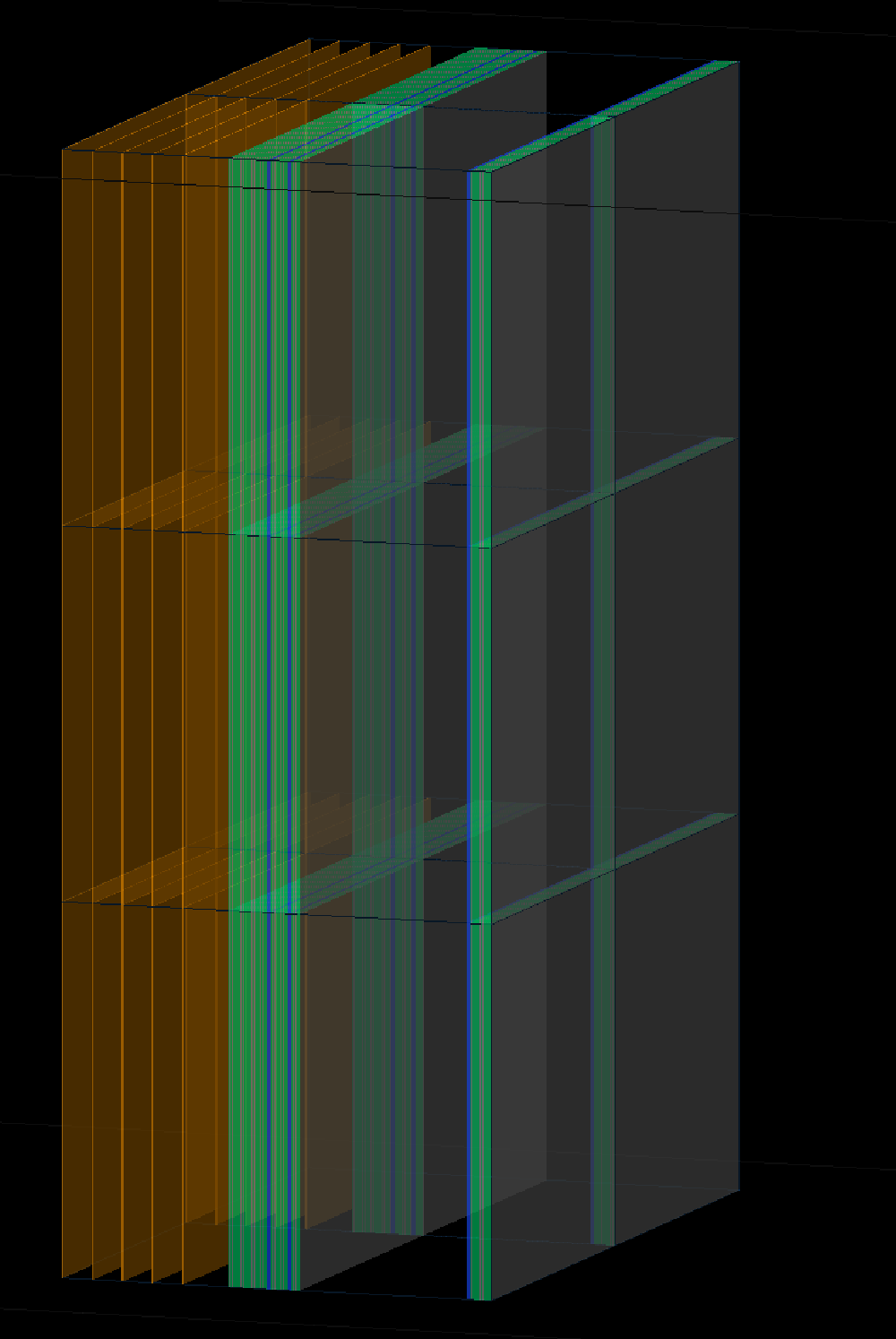}}
  \caption{Geometry of the calorimeter system: the lead ECAL with scintillator bar
  layers of alternating orientation, followed by the iron HCAL. The beam enters from the left.
  \label{fig:calo_geom}}
\end{figure}

The calorimeter system~\cite{Bonivento:2676562,Climescu:2025kdj}
(Fig.~\ref{fig:calo_geom}) consists of a lead electromagnetic section (ECAL)
of $20$ radiation lengths ($X_0$), sampled every $0.5\,X_0$ by scintillator
bars in alternating projections, with an air gap after the first $5\,X_0$,
followed by an iron hadronic section (HCAL) of five interaction lengths. It
spans $z=97.02$\,m--99.62\,m with a transverse face of
$4.32\times\SI{6.48}{m^2}$.

\subsection{Event samples}
\label{sec:samples}

The two stations lie at different distances from the target, so a common
transverse size would accept a different angular part of the neutrino beam. The samples for both detectors are therefore selected inside a common cone from the target,
of half-angle $\theta_{\max}=\rnd{2}{\ThetaMax}$~mrad, which keeps the near sample
inside the $40\times\SI{40}{cm^2}$ transverse size of the silicon-tungsten target: at the
downstream end of the near sample the cone has a half-width of
$\ConeRadiusWindowEnd$~cm. The production region is small compared with the
distances involved, and the simulated spectra in this matched acceptance agree
to within about $2\%$ in mean energy for all flavours
(Appendix~\ref{app:samples}).

At the near station both samples are taken from interactions with a vertex at
$\NueNearZmin<z<\NueNearZmax$~m, $\NueNearFrac\%$ of the SND interactions in
the cone: the upstream $\NearWindowPlates$ of the $\SNDTargetPlates$ tungsten
plates, which hold $\NearWindowTargetPct\%$ of the target interactions, and the
veto, with $\NearWindowVetoPct\%$ of the sample. For $\nue$ the granularity of
the target is needed for the separation from
NC-$\pizero$ events, and for $\numu$ the muons produced there are measured over
the full length of the magnetised iron.

At the far station
the samples are taken from interactions in the ECAL behind the air gap, to suppress the background from photons entering from upstream. Interactions in the HCAL are not used, since too
little instrumented material remains behind such a vertex. The
interaction-weighted baselines are $L_{\mathrm{near}}=\LNear$~m and
$L_{\mathrm{far}}=\LFar$~m, and the far-to-near ratio of CC interactions is
$R_{F/N}=\rnd{2}{\RFN}$ ($\numu$) and $\rnd{2}{\RFNE}$ ($\nue$). The struck nuclei are $\NearCompW\%$
tungsten at the near station (the rest: $\NearCompSi\%$ silicon and
$\NearCompC\%$ carbon), and $\rnd{0}{\FarCompPb}\%$ lead, $\rnd{0}{\FarCompC}\%$
carbon and $\rnd{0}{\FarCompAl}\%$ aluminium (from the high precision layers) at
the far station (the rest: $\FarCompH\%$ hydrogen); the
uncertainty this introduces in the far/near normalisation is discussed in
Sec.~\ref{sec:syst}.

\begin{figure*}[t]\centering
  \includegraphics[width=\textwidth]{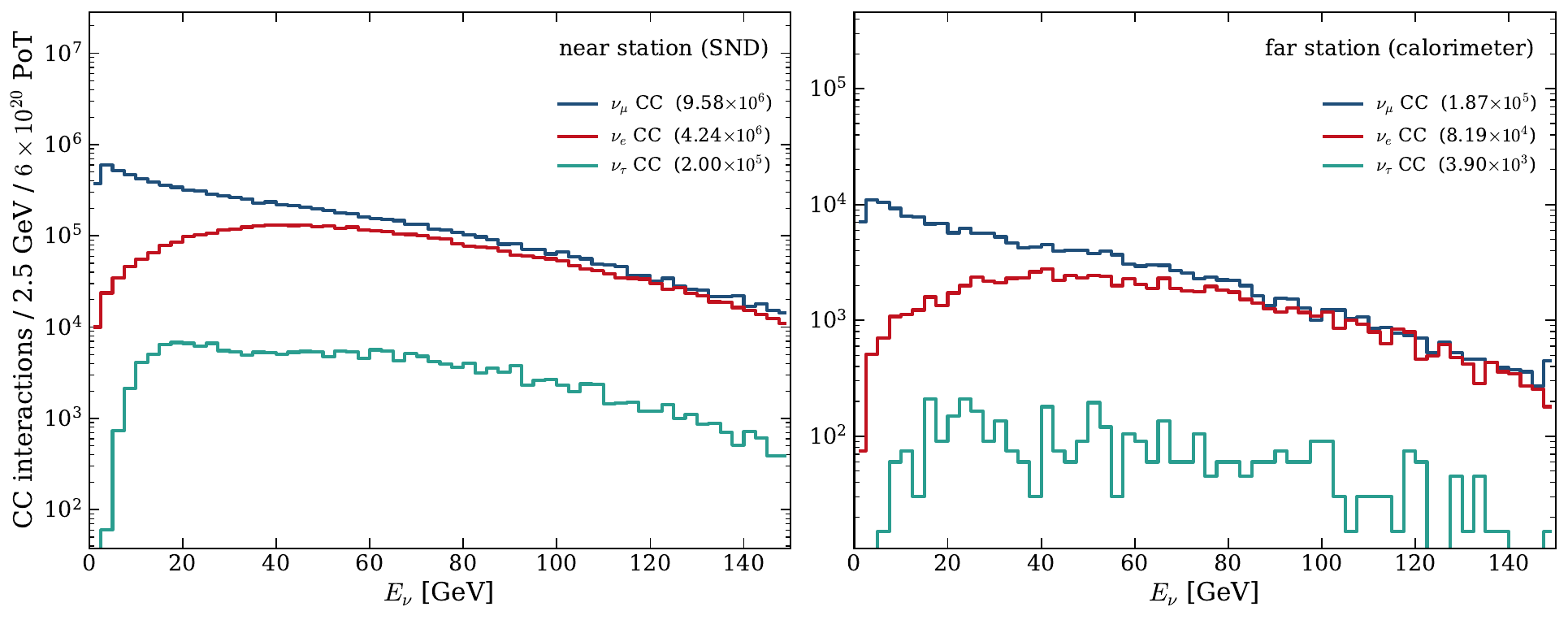}
  \caption{Energy spectra of CC-interacting neutrinos, by flavour, in the SND/near station
  (left) and in the calorimeter system/far station (right), in the matched acceptance, for
  $6\times10^{20}$~PoT.
  \label{fig:flux}}
\end{figure*}

The spectra of interacting neutrinos are shown in Fig.~\ref{fig:flux} and the
yields entering the fit in Table~\ref{tab:yields}. The $\numu$ flux originates
from pion, kaon and charm decays and the $\nue$ flux from kaon and charm
decays; the $\nutau$ flux comes only from $D_s\to\tau\nutau$ and the subsequent $\tau$
decay, and its CC rate is further reduced by the $\tau$ production threshold.

\input{table1.tex}

\section{Electron-neutrino reconstruction}
\label{sec:nue}

\subsection{Energy}
\label{sec:nueE}

\begin{figure*}[t]\centering
  \includegraphics[width=\textwidth]{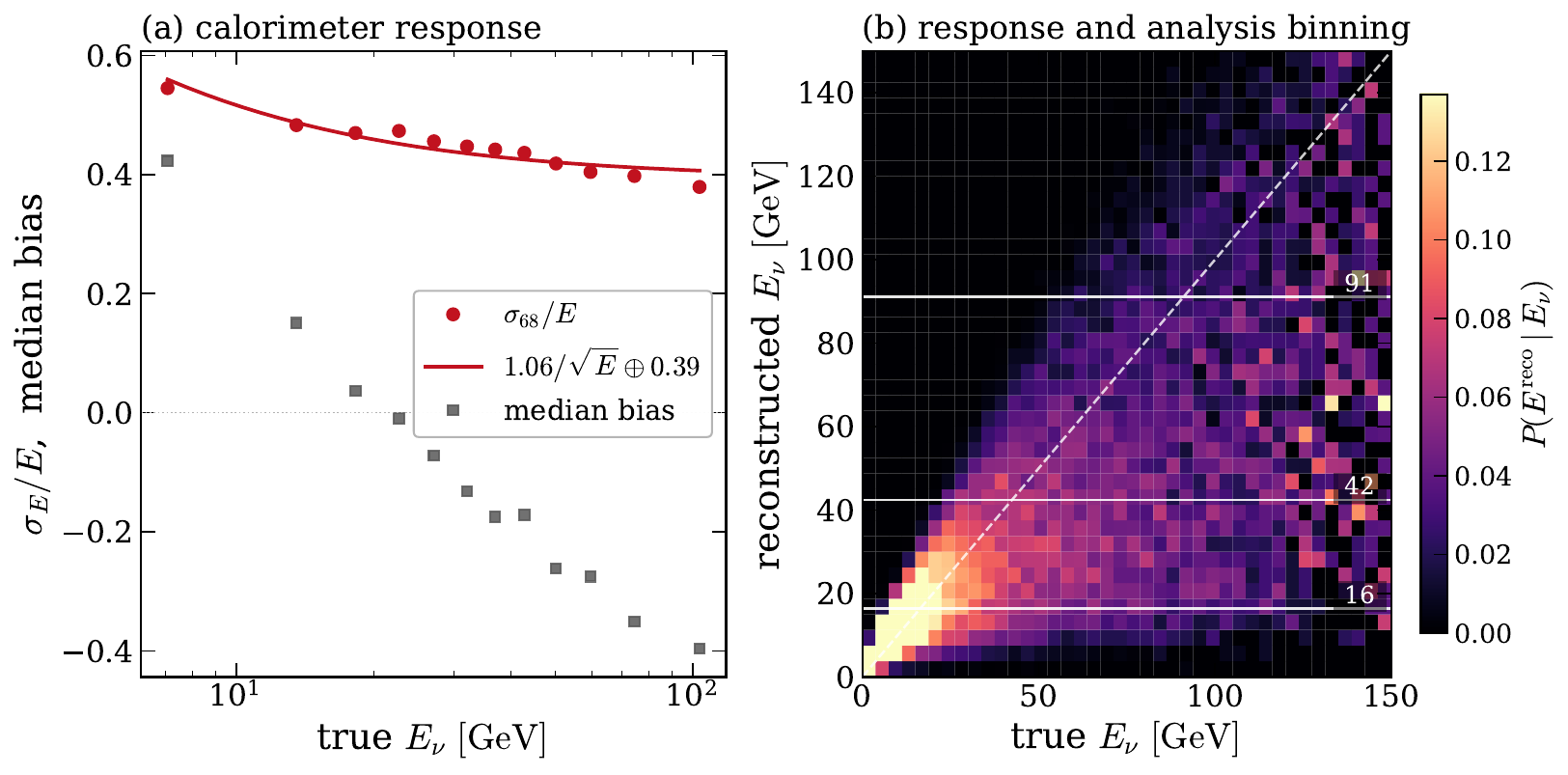}
  \caption{Electron-neutrino energy response at the far station. Left: $68\%$
  half-width of $(E_{\mathrm{reco}}-\Enu)/\Enu$ against true energy, with the
  fit of Eq.~\eqref{eq:calo_res}, and the median bias. Right: the
  true-to-reconstructed matrix with the analysis bin edges.
  \label{fig:nue_res}}
\end{figure*}

At the far station the reconstructed energy is a linear calibration of the
energy deposited in the ECAL, which contains the electron shower and the part
of the hadronic system contained there; it is a proxy for the neutrino energy,
not a measurement of the electron energy. Measured in bins of true energy, its
resolution is
\begin{equation}
\frac{\sigma_E}{E}\simeq\frac{1.06}{\sqrt{E/\mathrm{GeV}}}\oplus0.39 ,
\label{eq:calo_res}
\end{equation}
i.e.\ $0.42$ at \SI{50}{GeV} (Fig.~\ref{fig:nue_res}). The large constant term
reflects containment: the fluctuation of the escaping fraction does not decrease with
energy. The response also carries a median bias, from $+0.69$ at
\SIrange{2}{5}{GeV} to $-0.53$ above \SI{150}{GeV}: when $\sigma_E/E\sim0.5$ a
given deposit arises from a wide range of true energies, and an estimator
unbiased in bins of reconstructed energy is biased in bins of true energy. The
fit needs no unbiased estimator, only the correct response, which the migration
matrix built from the $27\,521$ simulated far $\nue$~CC interactions provides;
building it from the selected events only changes the limit by less than
$1\%$.

At the near station the energy observable is the sum of all deposits in the
silicon--tungsten target and in the fibre layers of the iron sections, which
collect the leakage of the shower; adding the scintillator layers degrades the
resolution, and they are not used. The observable is calibrated to the
neutrino energy on the $\NearNueCalibN$ simulated $\nue$~CC interactions of the
near sample. It is then unbiased, with $\sigma_{68}/E$ between
$\NearNueNuCalCoreLow$ and $\NearNueNuCalCoreHigh$ from $\NearNueNuCalEmin$ to
$\NearNueNuCalEmax$~GeV, and the near migration matrix is built from this
response.

\subsection{$\nue$\,CC versus NC-$\pizero$ separation}
\label{sec:nuesep}

The dominant background to $\nue$\,CC interactions is NC interactions producing a
$\pizero$, whose photons shower like the electron of a $\nue$~CC event. The electron shower starts at the vertex and carries the ionisation of a single
particle in its first layers, whereas the photons convert after a gap, deposit the ionisation of an $e^+e^-$ pair, and can form two separate showers. A gradient-boosted decision tree (\textsc{XGBoost}~\cite{XGBoost}) separates the
two, using the transverse occupancy image of the shower at the far station and
thirteen shower-shape variables at the near station, where an image-based
classifier overtrains on the available simulated statistics
(Appendix~\ref{app:reco}). All efficiencies and backgrounds in the fit are
evaluated on events not used to train the classifier that scores them.

\begin{figure}[t]\centering
  \includegraphics[width=\columnwidth]{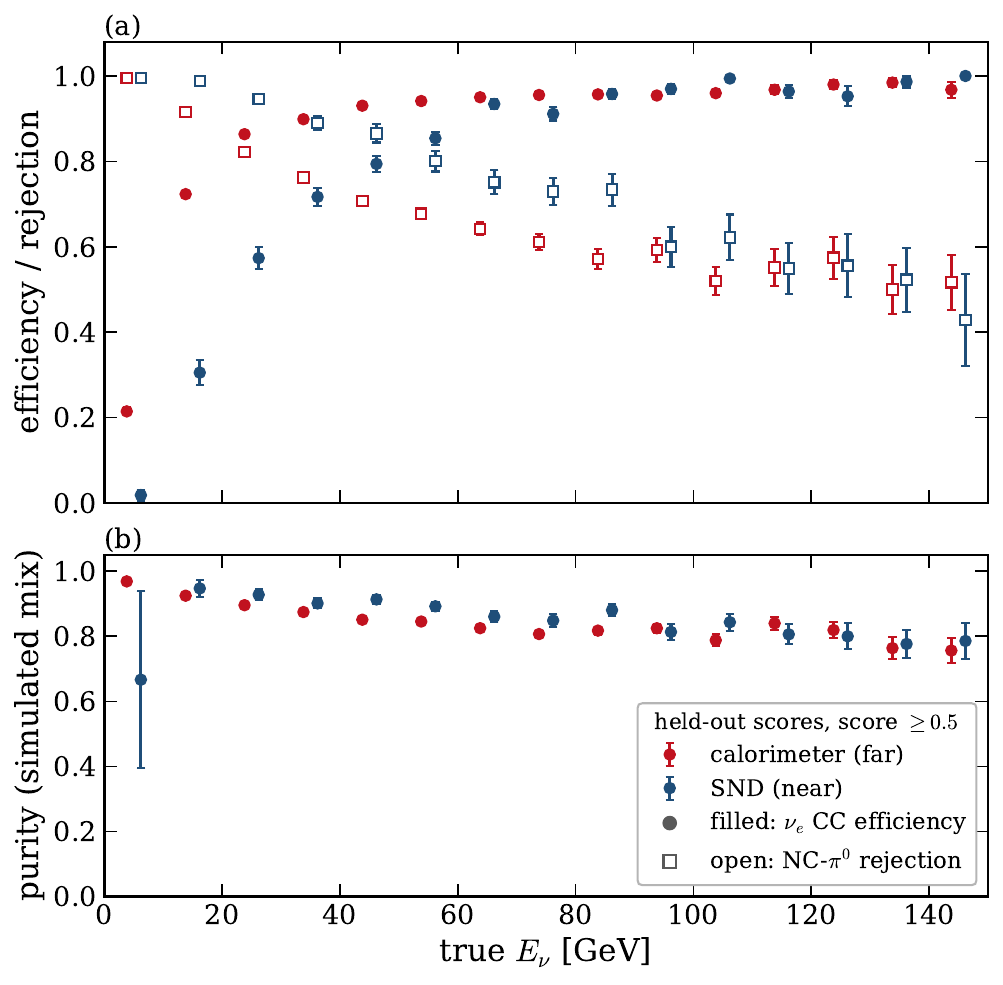}
  \caption{$\nue$~CC selection efficiency and NC-$\pizero$ rejection (upper) and
  purity of the simulated sample (lower) against true neutrino energy at score
  $\ge0.5$, for both stations, on events not used in training.
  \label{fig:id_eff}}
\end{figure}

At score $\ge0.5$ the classifier retains $\CaloSigEff\%$ ($\NueNearSigEff\%$)
of $\nue$~CC events and rejects $\CaloBkgRej\%$ ($\NueNearBkgRej\%$) of
NC-$\pizero$ events at the far (near) station, averaged over the simulated
samples. Weighted with the far $\nue$ spectrum, which is harder than that of the
simulated sample, the far efficiency is $90.2\%$, as in
Table~\ref{tab:yields}. Varying the
threshold between $0.3$ and $0.7$ at either station changes the $\nue$ limit by
less than $3\%$. The efficiency rises over the first few tens of GeV
(Fig.~\ref{fig:id_eff}), most slowly in the thin near target, where a soft
electron develops too little shower to be recognised. Since the oscillation
phase scales as $1/E$, this low-energy efficiency limits the $\nue$ reach at
low \dmsq{}.

\section{Muon-neutrino energy reconstruction}
\label{sec:numu}

The reconstructed energy of a $\numu$~CC interaction is
$\Ereco=E_\mu+E_{\mathrm{had}}$; the two stations measure the muon and hadronic
contributions differently.

\subsection{Far station}
\label{sec:numufar}

The calorimeter system has no magnetic field, and the muon momentum is estimated from
its multiple scattering: the deviations of the track from a straight line, its
kink angle and its depth in the HCAL are combined in a regression on the true
momentum (Appendix~\ref{app:reco}). Such an estimator is pulled towards the
typical momentum of its calibration sample: its median bias ranges from
$+\FarMuBiasTwoFive\%$ at \SIrange{2}{5}{GeV} to $\FarMuBiasFiftyHundred\%$ at
\SIrange{50}{100}{GeV}, although the core of the residual distribution is
narrow for penetrating muons ($\FarMuCoreFiftyHundred$--$\FarMuCoreTwentyFifty\%$
between 20 and \SI{100}{GeV}). The hadronic energy is measured
calorimetrically, with a resolution of $\FarHadRes\%$ and a bias in the energy
response within $\pm10\%$. The resulting far neutrino-energy resolution is $\FarNuMuERes\%$
($\FarNuMuECoreTwentyFifty\%$ at \SIrange{20}{50}{GeV}), and the muon accounts
for $\FarMuShare\%$ of its variance. The far migration matrix, built from
$17\,841$ simulated events, carries these biases in full.

\subsection{Near station}
\label{sec:numunear}

At the near station the muon momentum is measured from its deflection in the
magnetised iron. No field map of the muon-shield magnet is available yet: the
simulation is field-free, and the deflection is added analytically to the
simulated hits, which already carry the multiple scattering. A field
uncertainty acts as a momentum-scale uncertainty (Sec.~\ref{sec:syst}). Over
the \SI{1.47}{m} lever arm of a muon produced in the target the displacement
from its initial direction is \SI{6.2}{cm} at \SI{10}{GeV} and \SI{1.2}{cm} at
\SI{50}{GeV}, and bent tracks that leave the active area are lost: the field
removes $\FieldLossFracTen\%$ of the muons at \SI{10}{GeV},
$\FieldLossFracFifty\%$ at \SI{50}{GeV} and $\FieldLossAll\%$ overall. The
fit does not apply this loss separately: the muon acceptance and identification
are modelled by a single $90\%$ detection efficiency for every $\numu$~CC
interaction in the target, which is meant to include it, with full
identification of detected muons, wherever the muon stops or leaves the
detector; the same efficiency is assumed at the far station. Applying the
energy-dependent field loss explicitly instead changes the limit by less than
$1\%$. The resolution
is limited by multiple scattering in the iron, $\sigma_p/p\simeq18$--$20\%$
between 10 and \SI{100}{GeV}~\cite{Gluckstern}, and the near response is
parametric: the true muon momentum is smeared with this resolution, and the
true hadronic energy with an assumed calorimetric resolution of
$0.50/\sqrt{E/\mathrm{GeV}}\oplus0.05$ for the iron--scintillator sampling.

\section{Oscillation analysis}
\label{sec:osc}

\subsection{Two-baseline disappearance}
\label{sec:fold}

The oscillation probability, Eq.~\eqref{eq:posc}, is applied in true neutrino
energy before any detector effect. For each station $d$ the predicted content
of reconstructed-energy bin $i$ is
\begin{equation}
n^{d}_{i} =
M^{d}\sum_{j} \Phi^{d}_j\,\sigma^{d}_j\,\varepsilon^{d}_j\,
\big\langle P(\nu_\alpha\!\to\!\nu_\alpha)\big\rangle_{j,d}\,R^{d}_{ij},
\label{eq:folding}
\end{equation}
with the fiducial target mass $M^d$, the neutrino fluence $\Phi^d$ through the
station integrated over the exposure of $6\times10^{20}$~PoT, the CC cross
section per unit mass of the station's target $\sigma^d$, the selection
efficiency $\varepsilon^d$ and the migration
matrix $R^{d}_{ij}$ from true bin $j$ to reconstructed bin $i$; the
oscillation probability is averaged over each true-energy bin and over the
interaction-depth profile of the station. The neutrino source is treated as a
point at the start of the target; the production region, mostly within the
target, is short compared with both baselines, and folding an exponential
production-depth distribution of mean $\ProdZExpLambda$~m into the baselines
changes the limit at its minimum by less than $1\%$ in both channels. At the far station the
reconstructed bins are merged until each reaches a purity and efficiency of
$0.3$, so that the binning follows the resolution; this leaves six bins for
$\numu$ and four for $\nue$. Since the near station is itself \LNear~m from the
target, oscillations are already appreciable there: it is a second oscillation
measurement at a shorter baseline, not an unoscillated flux monitor.

The limits are quoted for one non-zero mixing element at a time
(Sec.~\ref{sec:model}), so that only disappearance contributes.

\subsection{Backgrounds}
\label{sec:bkg}

The selected NC-$\pizero$ and $\nutau\to e$ backgrounds are given in
Table~\ref{tab:yields}. Each selected NC-$\pizero$ event is
reconstructed with the $\nue$ energy estimator of its station applied to its
deposit, so that its response matrix, built from the simulated NC-$\pizero$
events, maps the energy of the parent neutrino directly to the reconstructed
energy. Its component from $\nue$ parents disappears with $P(\nue\to\nue)$,
since a neutrino that has oscillated into the sterile state does not interact;
the components from $\numu$ and $\nutau$ parents do not. The $\nutau$~CC
interactions with $\tau\to e$ cannot be
distinguished from the signal and are included as a non-oscillating component.

\subsection{Systematic uncertainties}
\label{sec:syst}

Uncertainties in the overall normalisation common to the two stations, from
the number of protons on target, the neutrino production in the target and the
cross-section scale, cancel in the near/far comparison; this common
normalisation is left free in the fit. The remaining
uncertainties and their effect on the limits are summarised in
Table~\ref{tab:systsum}.

\emph{Flux shape.} The flux uncertainty is a covariance in true neutrino energy,
\begin{equation}
C_{jj'} = \sigma_\Phi^2\,
\exp\!\left[-\frac{\ln^2(E_j/E_{j'})}{2\ell^2}\right],
\label{eq:fluxcov}
\end{equation}
propagated through the response of both stations and diagonalised, each
principal component becoming one nuisance parameter~\cite{DUNE_TDR,T2K_2017}.
The near spectrum constrains this common distortion, and only the part the two
stations cannot resolve remains degenerate with an oscillation. We take
$\sigma_\Phi=20\%$, the level of the discrepancy between the measured muon flux
from a replica SHiP target and the simulation~\cite{SHiP:2020hyy}, and
$\ell=0.5$ in $\ln E$; both are assumptions, since that measurement does not
determine the neutrino-energy covariance. Below $\dmsq\simeq10^3$~\si{eV^2} the
sensitivity changes by about $10\%$ for correlation lengths down to $0.05$;
above it only the near station contributes and the reach depends on $\ell$.
The distortion is assumed common to the two stations; a component that differs
between them, for instance from the extended production region, is not
evaluated here. The covariance acts on the summed $\numu+\bar\nu_\mu$ flux, so
the shapes of the two are taken as fully correlated, as expected from their
common hadroproduction origin; if they were independent, the $\numu$ limit
would weaken by up to about $\FluxIndepPct\%$, a dependence that the muon-charge
measurement at the near station removes, reproducing the quoted limit to within
$\NearChargeDevPct\%$.

\emph{Far/near normalisation.} One nuisance parameter with a Gaussian prior $\srel$
scales the far rate relative to the near rate. It covers the ratio of cross
sections per unit mass in the two targets, the target masses and the relative
selection efficiency of the two detector technologies. The target composition
is part of the simulated yields; $\srel$ covers the uncertainty on the ratio.
Between tungsten and lead, charged-lepton data show nuclear modifications
scaling as $A^\alpha$ with $|\alpha|<0.05$~\cite{E139}, i.e.\ differences below
$0.6\%$ at any Bjorken $x$; about $30\%$ of the far interactions occur on carbon and
aluminium, for which model predictions of cross-section ratios to heavy nuclei
can fail at the $10\%$ level~\cite{MINERvA2014}. We adopt $\srel=1\%$ as a
benchmark performance requirement for the relative near/far normalisation,
not as a precision already demonstrated; it is tighter than
what experiments with dissimilar near and far detectors have achieved so
far~\cite{T2K_2017}. With $\srel=2\%$ the limits become $\Umufour=\UmuLimitRelTwo$
and $\Uefour=\UeLimitRelTwo$ (Table~\ref{tab:systsum}). The $\nue$ channel,
whose far spectrum has only four energy bins, draws its sensitivity mostly from
the ratio of the event rates at the two stations, which a far/near
normalisation change mimics directly, and so depends strongly on this
assumption. The $\numu$ channel draws its sensitivity from the energy
dependence of the deficit, which a normalisation change cannot reproduce, and
is almost unaffected.

\emph{Backgrounds.} The NC-$\pizero$ and $\nutau\to e$ backgrounds have
independent normalisations at each station, with Gaussian priors of $30\%$ and
$50\%$, the latter reflecting the uncertainty on $D_s$ production.

\emph{Detector response.} The limits assume that the energy scales and responses
of the two stations are known. Independent uncertainties on the muon and
hadronic energy scales of each station, or an uncorrelated bin-to-bin
uncertainty on the response shape, weaken the $\numu$ sensitivity
(Table~\ref{tab:systsum}), the largest contribution coming from the
far-station muon energy. The finite simulated statistics contribute about
$1\%$.

\begin{table}[t]
\centering
\caption{Effect of variations of the analysis assumptions on the $95\%$~CL
median limit at its minimum, as a factor relative to the quoted results
($\Umufour=\UmuLimit$, $\Uefour=\UeLimit$).}
\label{tab:systsum}
\begin{tabular}{lcc}
\toprule
Variation & $\numu$ & $\nue$ \\
\midrule
Flux correlation length $\ell=0.05$ & $+\LSpreadPct\%$ & $+\LSpreadPctE\%$ \\
Far/near normalisation, $2\%$ & $\times\SigRelTwoMu$ & $\times\SigRelTwoE$ \\
Far/near normalisation, $5\%$ & $\times\SigRelFiveMu$ & $\times\SigRelFiveE$ \\
Energy scales, $0.5\%$ each & $\times\ScaleCompHalfMu$ & -- \\
Energy scales, $1\%$ each & $\times\ScaleCompOneMu$ & -- \\
Response shape, $1\%$ per bin & $\times\ShapeOnePctMu$ & $\times\ShapeOnePctE$ \\
Simulated statistics (r.m.s.) & $\MCStatMu\%$ & $\MCStatE\%$ \\
\bottomrule
\end{tabular}
\end{table}

\begin{figure*}[t]\centering
  \includegraphics[width=0.49\textwidth]{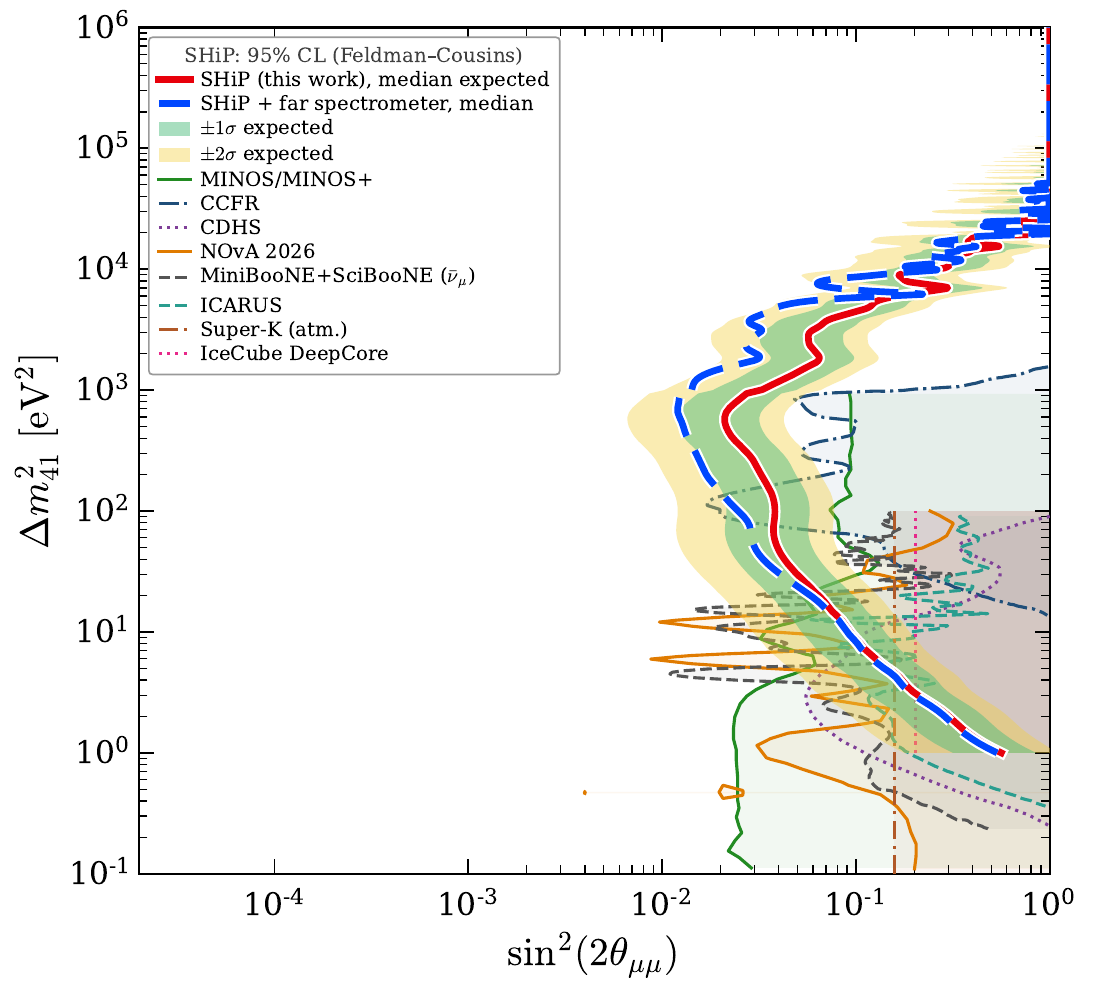}\hfill
  \includegraphics[width=0.49\textwidth]{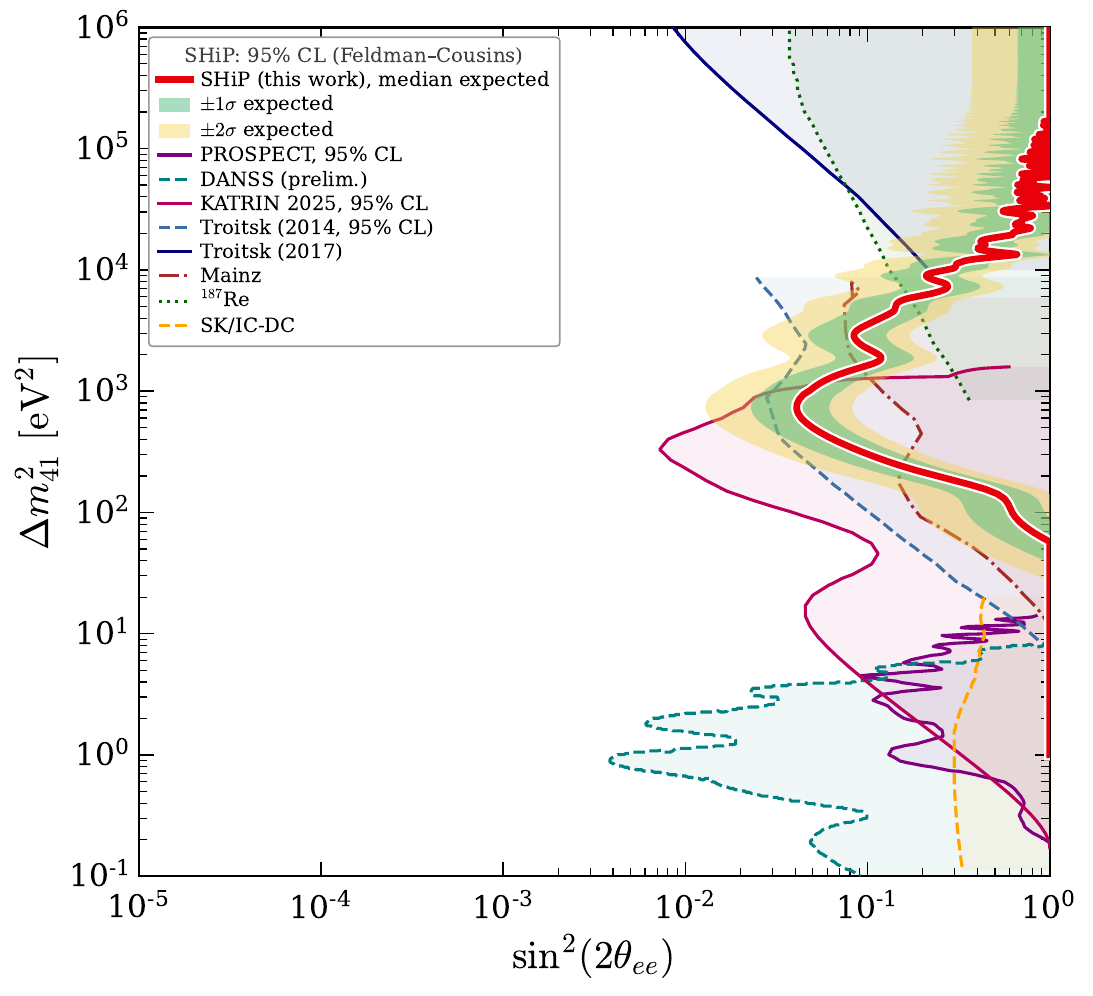}
  \caption{Projected SHiP $95\%$~CL median expected sensitivity, with the
  $\pm1\sigma$ and $\pm2\sigma$ bands, for $6\times10^{20}$~PoT. Left:
  $\sstt_{\mu\mu}$, with accelerator
  limits~\cite{CDHSW,CCFR,MINOS_sterile,MBSB2012,NOvA2026,ICARUS:2026} and
  atmospheric limits~\cite{SuperK2015,IceCubeDC2024}, each drawn over the
  \dmsq{} range of its publication; the dashed line is
  the median sensitivity with an idealised far muon spectrometer
  (Sec.~\ref{sec:sens}). Right: $\sstt_{ee}$, with reactor
  (PROSPECT~\cite{PROSPECT}, DANSS~\cite{DANSS:2025}), $\beta$-decay
  (KATRIN~\cite{KATRIN_sterile_KNM2,KATRIN:2025lph},
  Troitsk~\cite{Troitsk2014,Troitsk2017}, Mainz~\cite{Mainz2013},
  $^{187}$Re~\cite{PhysRevLett.86.1978}) and atmospheric~\cite{Dentler:2018sju}
  constraints. Below $\dmsq\approx10^3$~\si{eV^2} both stations contribute and
  the curves are stable to $\approx10\%$ under changes of the flux-correlation
  model; above it only the near station contributes, the curves use the most
  conservative flux model ($\ell=0.05$), and the reach depends on that model;
  the narrow features of the curves there, a few per cent wide in \dmsq{}, are
  numerically converged and arise where the near-station oscillation pattern can
  be absorbed by the flux-shape model at the near-station energy resolution, so
  they are not resolvable features of the sensitivity. External limits are
  shown at their published confidence level: $95\%$ where labelled, $90\%$
  otherwise. The SHiP curves start at $\dmsq=1$~\si{eV^2}.
  \label{fig:osc}}
\end{figure*}

\begin{figure*}[t]\centering
  \includegraphics[width=0.49\textwidth]{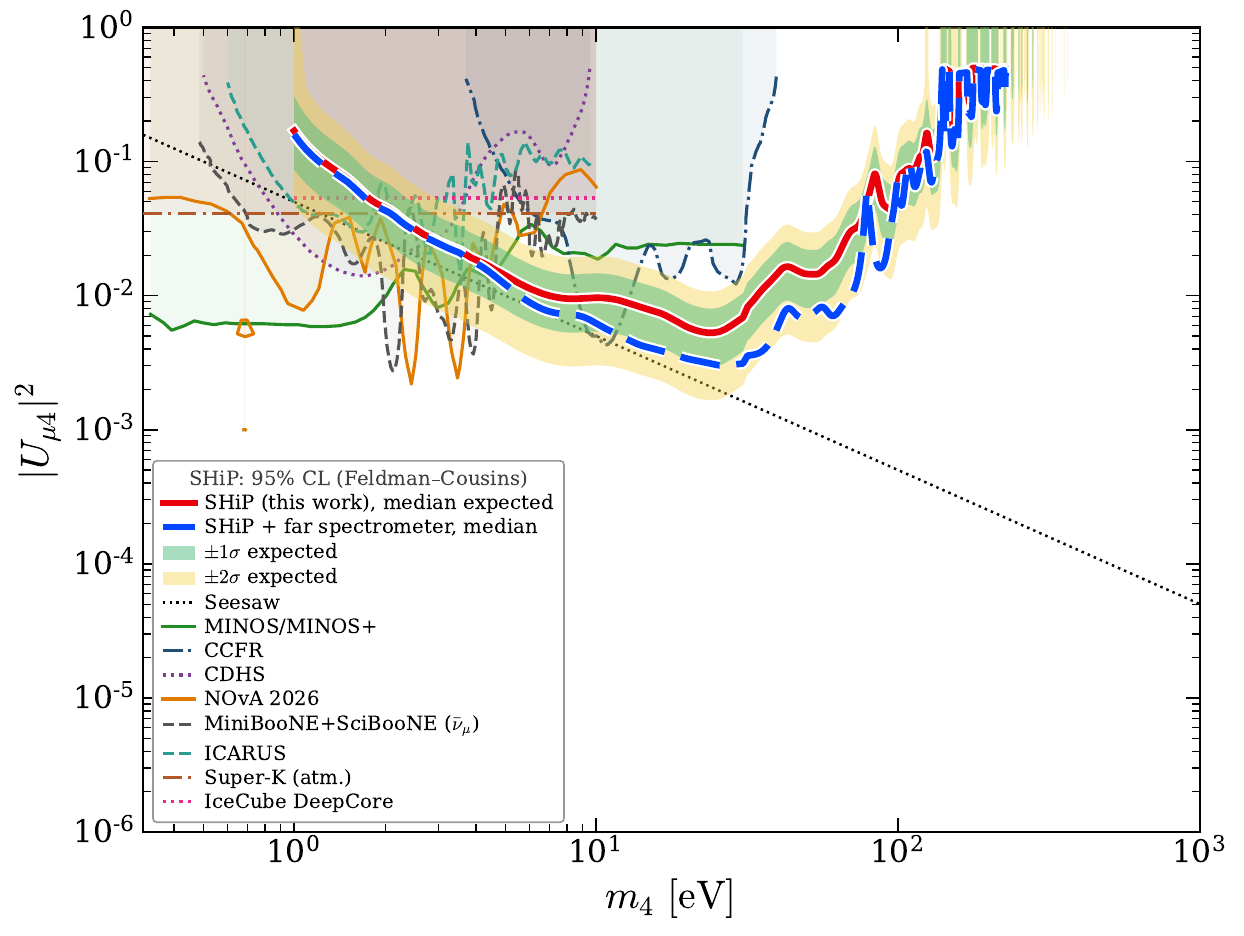}\hfill
  \includegraphics[width=0.49\textwidth]{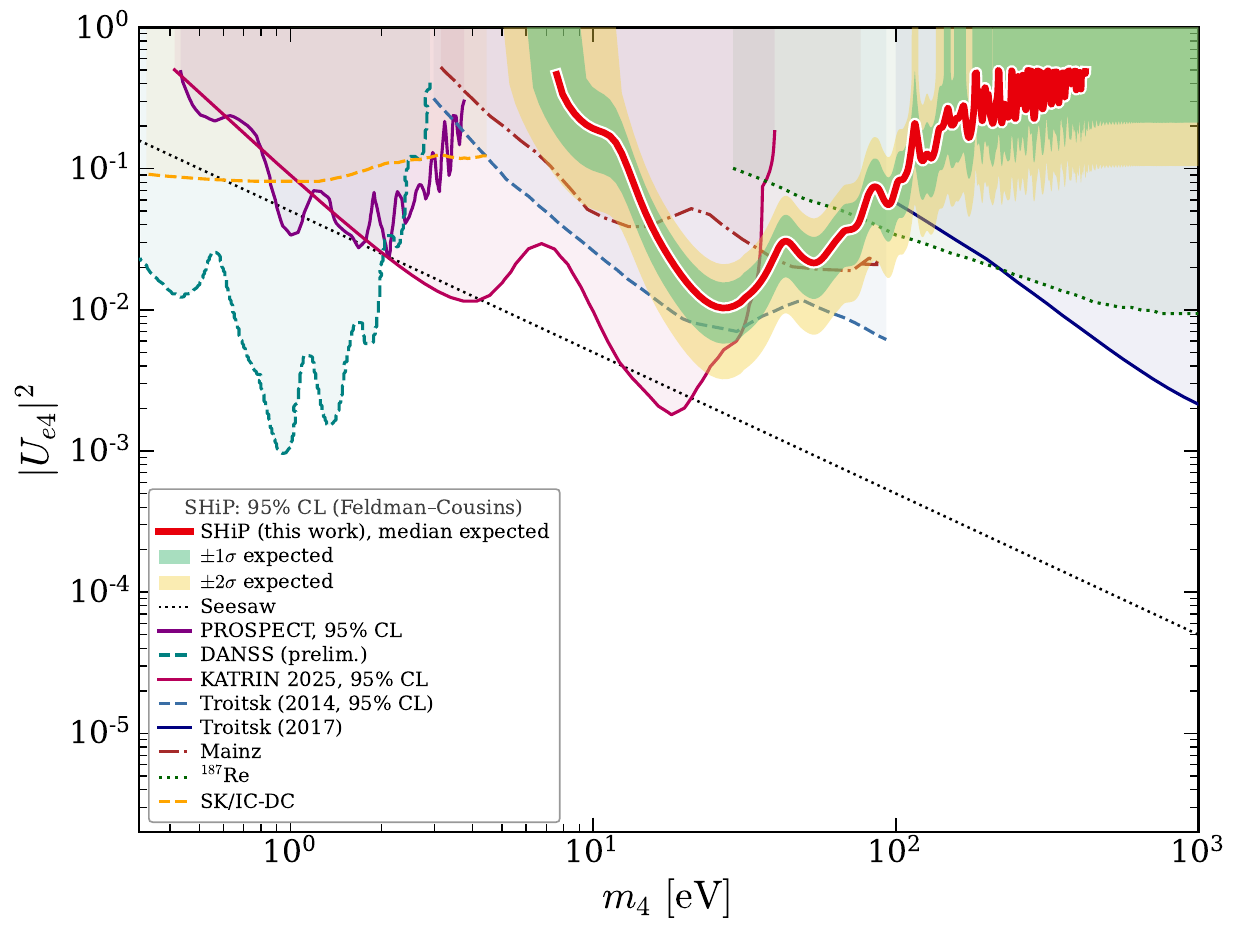}
  \caption{The sensitivities of Fig.~\ref{fig:osc} in the $(\mfour,\Umufour)$ (left)
  and $(\mfour,\Uefour)$ (right) planes, with the $\pm1\sigma$ and $\pm2\sigma$
  bands, existing limits and the seesaw line. The remarks of Fig.~\ref{fig:osc} on
  $\dmsq\gtrsim10^3$~\si{eV^2} ($\mfour\gtrsim\SI{30}{eV}$) and on the confidence
  levels apply; in particular, the SHiP curves start at $\mfour=\SI{1}{eV}$
  ($\dmsq=\SI{1}{eV^2}$).
  \label{fig:hnl}}
\end{figure*}

\subsection{Statistical procedure}
\label{sec:stat}

The near and far spectra are compared with the no-oscillation expectation in a
Gaussian approximation to the Poisson likelihood, with the nuisance parameters
profiled analytically. The $95\%$~CL median expected limit and its bands follow
from a Feldman--Cousins construction~\cite{FeldmanCousins} for an amplitude
estimator bounded at \mbox{$\sstt\ge0$}. Pseudo-experiments reproduce the analytic
limit to within $2\%$ ($\numu$) and $4\%$ ($\nue$), and
injected signals are recovered in both channels (Appendix~\ref{app:stat}).

Neither station is redundant: the calorimeter system alone reaches at best
$\Umufour\simeq\FarOnlyMinMu$, a factor of about $\rnd{0}{\NearWorthMu}$ above
the combined result (about $\rnd{0}{\NearWorthE}$ for $\nue$). With the flux shape assumed known, the near
station alone would nearly reproduce the combined result; with the flux-shape
uncertainty, a deficit at one station is equally well explained by a
mismodelled spectrum, and only the comparison of the two baselines separates
the two.


\section{Sensitivity}
\label{sec:sens}

The projected $95\%$~CL median sensitivities are shown in Fig.~\ref{fig:osc}.
The $\numu$ channel is strongest where the oscillation phases at the two
stations differ most, reaching $\Umufour\simeq\UmuLimit$ at
$\dmsq\simeq\UmuLimitDm$~\si{eV^2} ($\mfour\simeq\UmuLimitMass$~eV). Its median
sensitivity lies below the existing $\numu$-disappearance limits for
$\DnineLow\lesssim\mfour\lesssim\DnineHigh$~eV, where both stations contribute,
and remains below them up to $\mfour\simeq\DnineNearOnlyHigh$~eV with the near
station alone in every flux-correlation model considered; it is also below them
in a narrow window at \mbox{$\DnineWindowLow$--$\DnineWindowHigh$~eV}. With an
additional $1\%$ response-shape uncertainty it lies below them over
\mbox{$\DnineShapeSpanLow$--$\DnineShapeSpanHigh$~eV}, except in two narrow intervals
near $\DnineShapeGapOne$ and $\DnineShapeGapTwo$~eV, where CCFR is stronger by
up to $\DnineShapeGapDeficitPct\%$. These statements compare a median
expected sensitivity at $95\%$~CL with observed limits, most of which are given
at $90\%$~CL; the comparison is therefore conservative.

\emph{Far muon spectrometer.} The least precise ingredient of the muon channel
is the far muon momentum (Sec.~\ref{sec:numufar}). In an idealised study, a
magnetised spectrometer behind the calorimeter system measuring the momentum of the
muons that cross it to $\SpecResolution\%$ improves the far $\numu$ energy
resolution from $\SpecFarResCalo\%$ to $\SpecFarResSpec\%$ and the sensitivity
to $\Umufour\simeq\UmuLimitSpecFine$ (dashed line in Fig.~\ref{fig:osc}). It
also reduces the cost of the energy-scale uncertainties from a factor
$\ScaleCompOneMu$ to $\SpecScaleCostMu$ and lets the far station contribute
above $\dmsq\approx10^3$~\si{eV^2}. Beyond this point the sensitivity is
limited by the flux-shape and far/near normalisation uncertainties rather than
by resolution. With the muon charge measured at both stations, the
spectrometer would also allow the disappearance amplitudes of neutrinos and
antineutrinos to be fitted independently, without assuming CPT symmetry; the
projected sensitivities to $\Umufour$ are then $\UmuLimitSpecNu$ from the
$\numu$ amplitude and $\UmuLimitSpecNubar$ from the $\bar\nu_\mu$ amplitude.
Such a measurement could use a spectrometer of the NA62 type: the NA62
spectrometer~\cite{NA62:2017rwk} has been dismantled and is kept in the ECN3
hall.

The $\nue$ channel reaches $\Uefour\simeq\UeLimit$ at
$\dmsq\simeq\UeLimitDm$~\si{eV^2} ($\mfour\simeq\UeLimitMass$~eV). In this mass
range the tritium $\beta$-decay searches constrain the electron mixing more
strongly wherever they apply: KATRIN~\cite{KATRIN:2025lph,KATRIN_sterile_KNM2}
below $\mfour\approx\SI{35}{eV}$, and Troitsk up to $\mfour\approx\SI{90}{eV}$
(2014~\cite{Troitsk2014}) and $\approx\SI{400}{eV}$ (2017~\cite{Troitsk2017});
$^{187}$Re decay~\cite{PhysRevLett.86.1978} covers higher masses. The SHiP measurement is
obtained by a different method, an oscillation over two known baselines rather
than a distortion of a $\beta$ spectrum, with different systematic
uncertainties.

Figure~\ref{fig:hnl} shows the same sensitivities in the mass--mixing plane
(Sec.~\ref{sec:model}), complementary to the decay searches of the SHiP
programme at the GeV scale.

\emph{Comparison with earlier studies.} Reference~\cite{Choi2025} reaches
$|U_{\alpha4}|^2\approx0.1$ near $\dmsq\approx10^3$~\si{eV^2} with a near
detector alone and the $10^{-2}$ level with an additional far detector, at
$90\%$~CL for $2\times10^{20}$~PoT. The two studies differ in assumptions that
act in opposite directions: that study assumes a far detector with the same
acceptance as the near one, whereas the calorimeter system collects about ten
times fewer interactions than the SND target ($R_{F/N}$, Sec.~\ref{sec:samples});
it uses a three times smaller exposure; and it includes normalisation
uncertainties only, whereas the present analysis also includes an
energy-dependent flux-shape uncertainty, its leading systematic effect
(without it the $\numu$ limit would be $\Umufour\simeq\UmuLimitNoShape$), and
quotes $95\%$~CL. With these differences, the two studies reach mixings of the
same order.

\section{Conclusions and outlook}
\label{sec:concl}

The SHiP SND and calorimeter system form a near/far pair at
$\LNear$ and $\LFar$~m from the target. With responses derived from simulation,
$\nue$~CC events are reconstructed with a containment-limited resolution of
$1.06/\sqrt{E/\mathrm{GeV}}\oplus0.39$ in the calorimeter system and separated from
NC-$\pizero$ events with $\CaloSigEff\%$ efficiency at $\CaloBkgRej\%$
rejection, and with $\NueNearSigEff\%$ at $\NueNearBkgRej\%$ in the SND target;
the muon momentum is measured from magnetic deflection in the SND with a
resolution of about $20\%$, and estimated from multiple scattering in the
calorimeter system. The projected $95\%$~CL sensitivities are $\Umufour\simeq\UmuLimit$
and $\Uefour\simeq\UeLimit$ near $\mfour\simeq\UmuLimitMass$ and
$\UeLimitMass$~eV; in the muon channel this lies below existing limits for
$\DnineLow\lesssim\mfour\lesssim\DnineHigh$~eV. The result assumes known energy
scales and responses of the two stations and a $1\%$ far/near normalisation;
the relative calibration of the two stations' energy scales is the main
experimental requirement for the muon channel.

The SHiP detector design can still be optimised for this
measurement: its least precise ingredient is the far muon momentum, and a muon
spectrometer behind the calorimeter system would strengthen the muon channel and
reduce its dependence on the energy scales. Other detector options have been discussed for the facility,
among them a liquid-argon time-projection chamber~\cite{Ferrillo:2023hhg}.
This work demonstrates that the baseline SHiP detector configuration can conduct
competitive studies of sterile-neutrino-induced oscillations.

\begin{acknowledgments}
We thank Nick van Remortel and Alfons Weber for their comments.
M.~Climescu has received support from the FWO, under grant no.\ 12A4O26N
(Belgium). This work is supported by the Swiss National Science Foundation through an Ambizione grant No. PZ00-2\_240997.
\end{acknowledgments}

\FloatBarrier
\appendix

\begin{figure*}[!t]\centering
  \includegraphics[width=\textwidth]{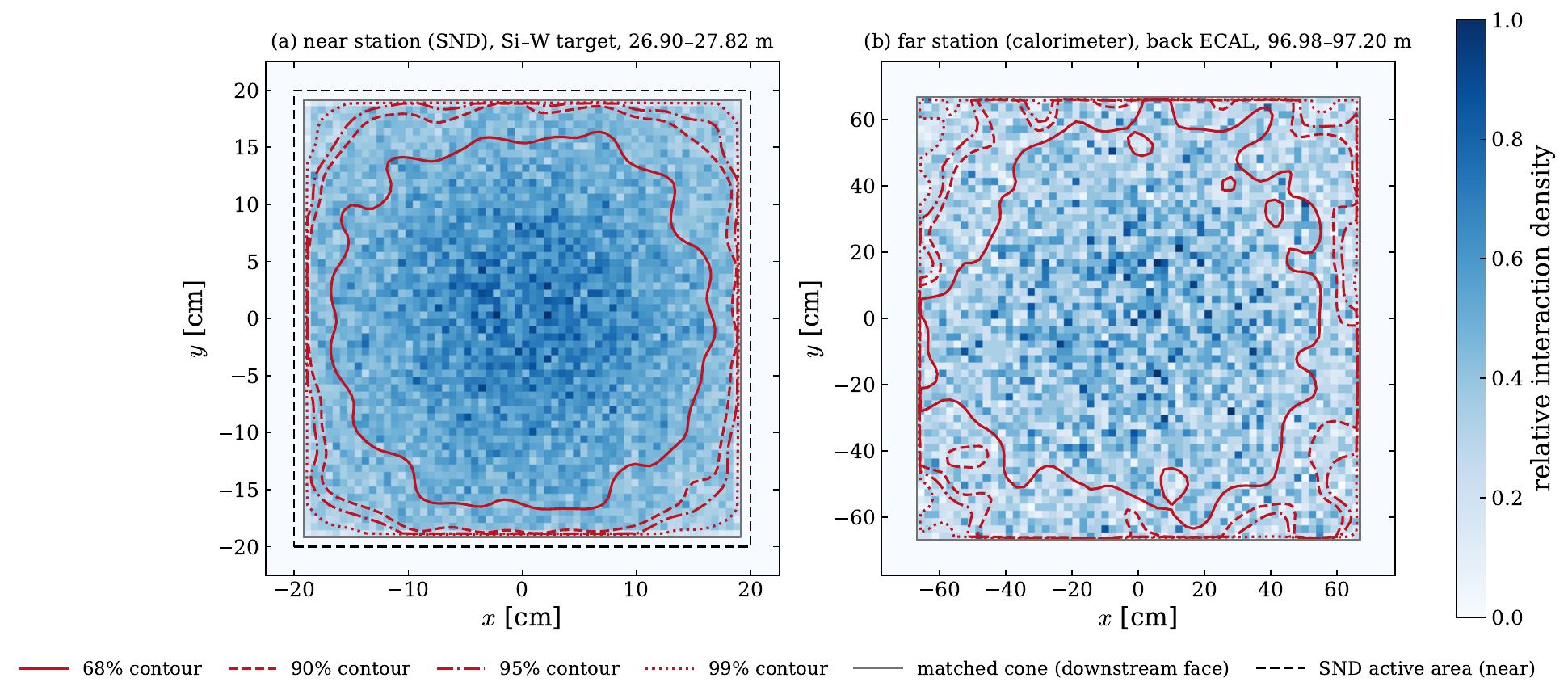}
  \caption{Transverse distribution of CC interaction vertices in the near
  (silicon--tungsten target) and far (ECAL) fiducial volumes, with the
  matched cone and the $68$, $90$, $95$ and $99\%$ containment contours.
  \label{fig:vtx}}
\end{figure*}

\begin{figure*}[!t]\centering
  \includegraphics[width=0.7\textwidth]{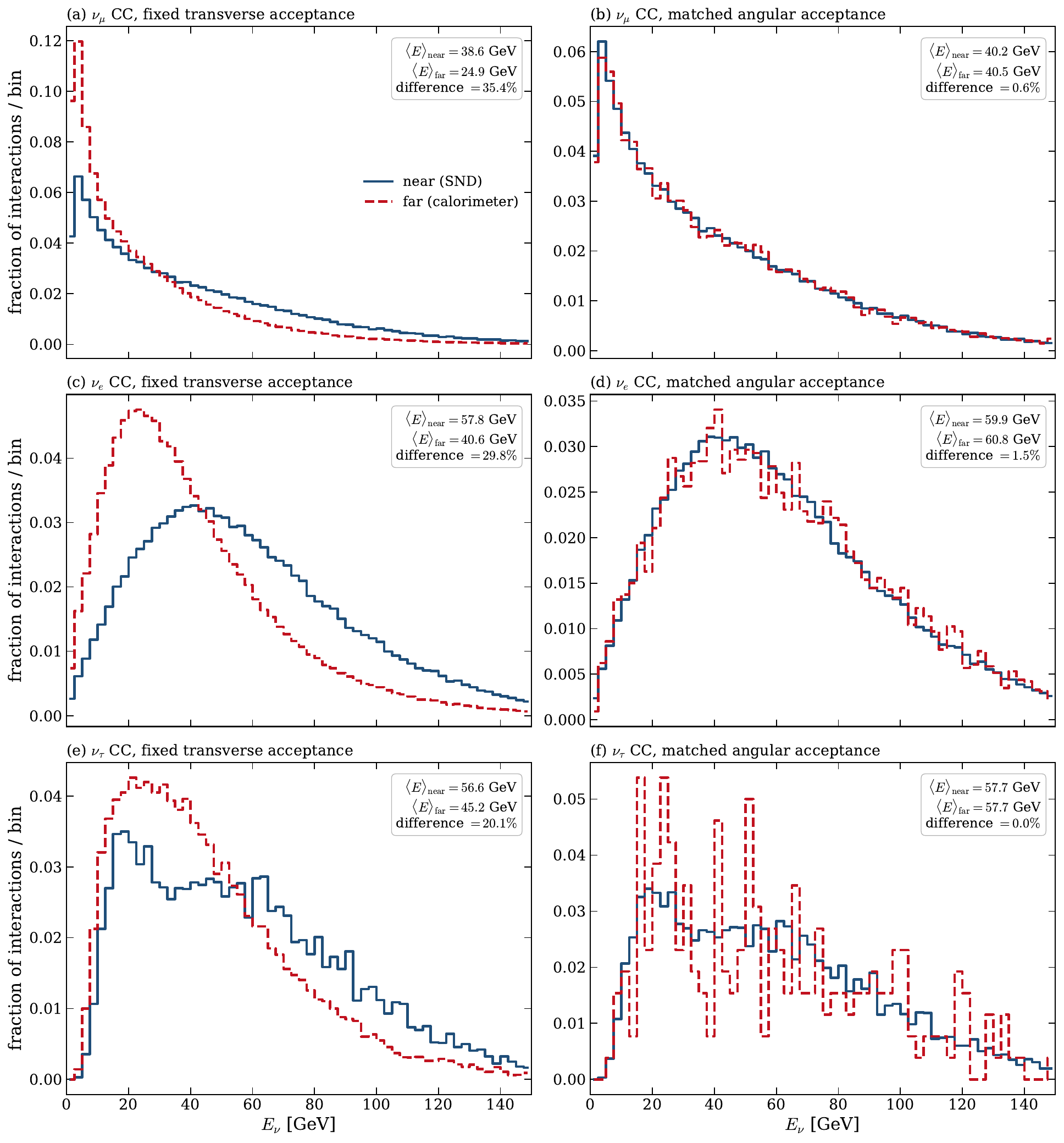}
  \caption{CC interaction spectra of $\numu$, $\nue$ and $\nutau$ at the two
  stations, normalised to unit area, with a common transverse acceptance, for
  which the two stations sample different parts of the beam, and with the
  matched cone, for which the spectra coincide.
  \label{fig:matched_all}}
\end{figure*}

\begin{figure*}[!t]\centering
  \includegraphics[width=0.49\textwidth]{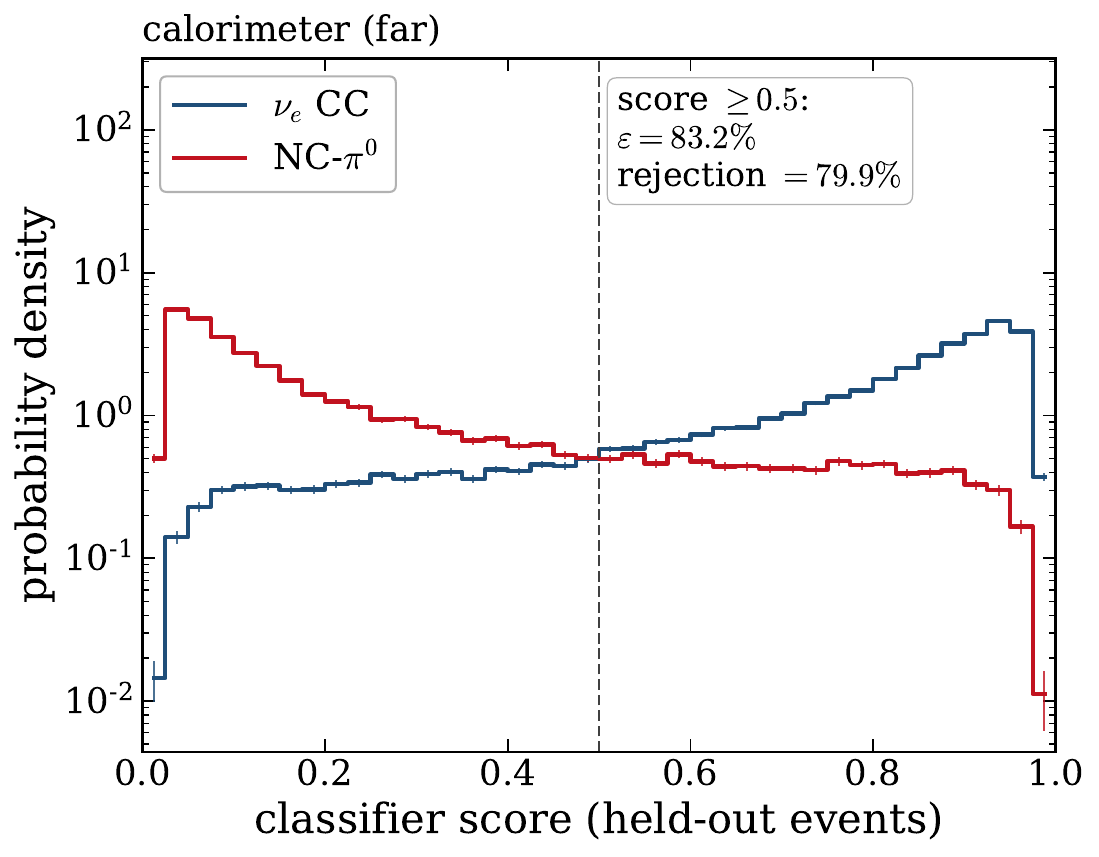}\hfill
  \includegraphics[width=0.49\textwidth]{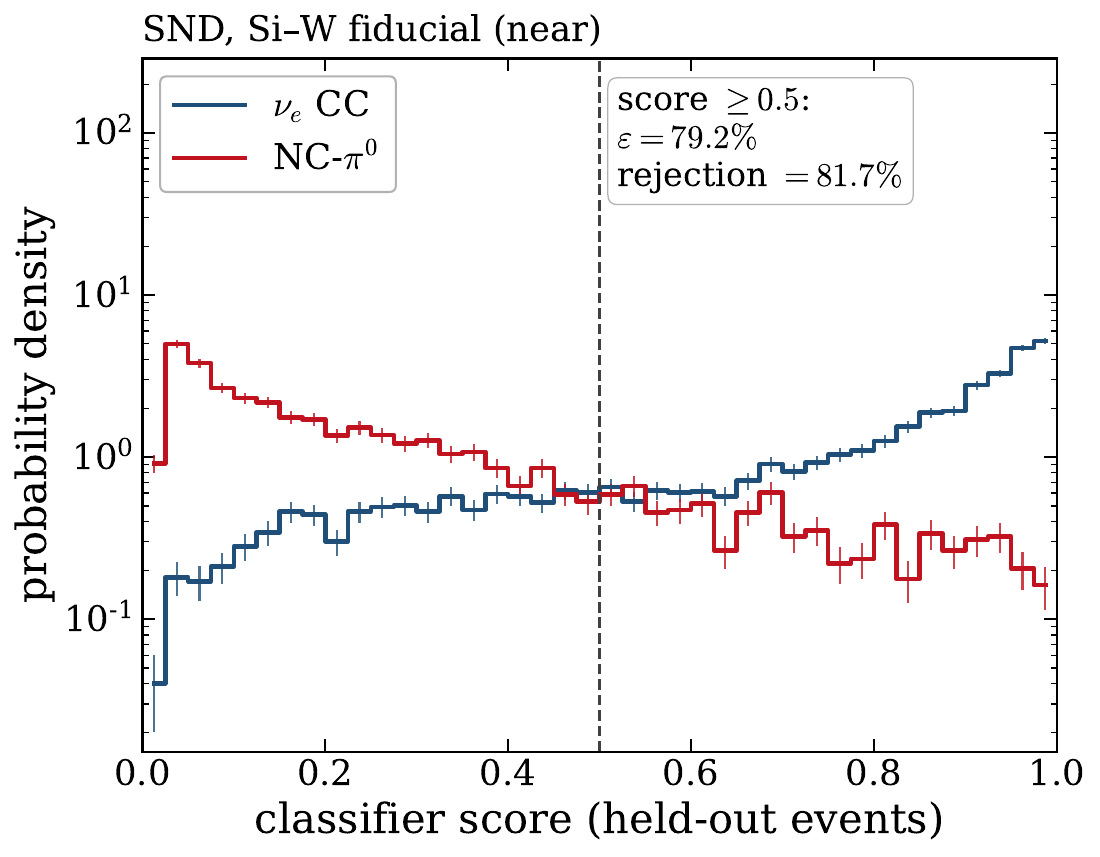}
  \caption{Classifier output for $\nue$~CC and NC-$\pizero$ events at the far
  (left) and near (right) station, on events not used in training, with the
  working point, and the efficiency and rejection against threshold.
  \label{fig:bdt_score}}
\end{figure*}

\begin{figure*}[!t]\centering
  \includegraphics[width=\textwidth]{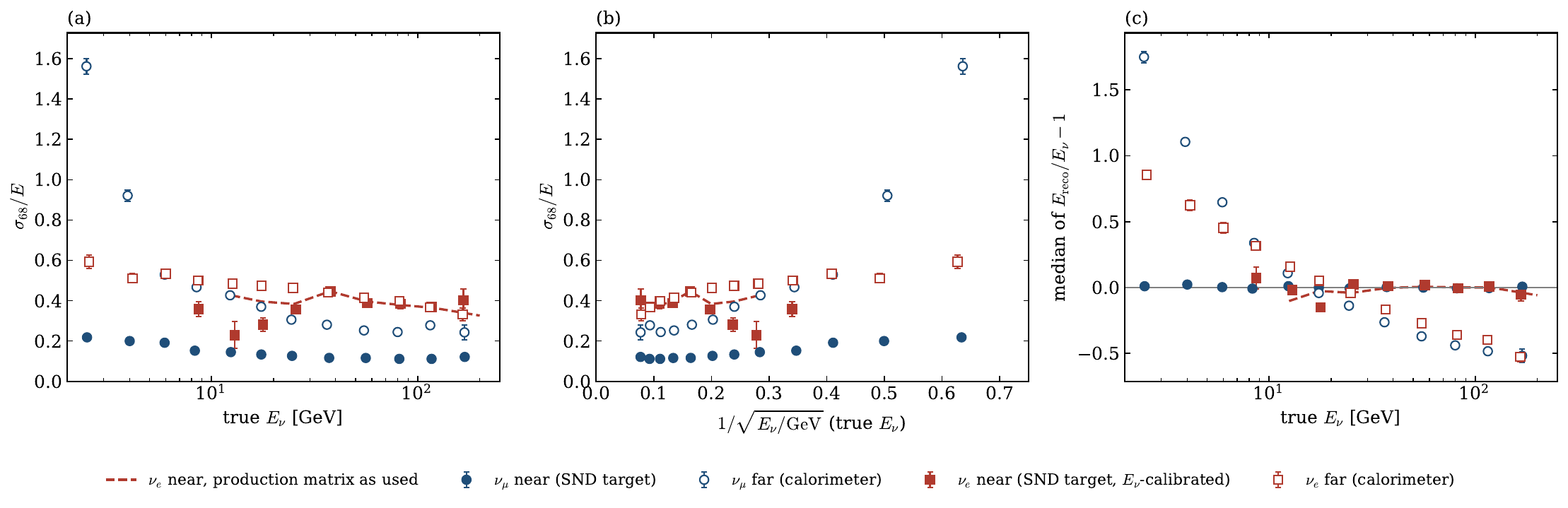}
  \caption{Neutrino-energy response of the four samples as used in the fit:
  $\sigma_{68}/E$ of the reconstructed neutrino energy against the true $\Enu$
  (a) and against $1/\sqrt{\Enu/\mathrm{GeV}}$ (b), and the median bias (c).
  Near $\numu$: parametric response (Gluckstern muon $\oplus$
  $0.50/\sqrt{E}\oplus0.05$ hadrons); far $\numu$ and $\nue$: simulated events,
  with statistical errors; near $\nue$: the simulated response of the
  estimator calibrated to the neutrino energy (markers, with bootstrap
  errors) and the migration matrix built from it (dashed), drawn above
  \SI{10}{GeV}, where its \SI{7.5}{GeV} reconstructed-energy bins resolve the
  response.
  \label{fig:nu_eres}}
\end{figure*}

\begin{figure*}[!t]\centering
  \includegraphics[width=0.49\textwidth]{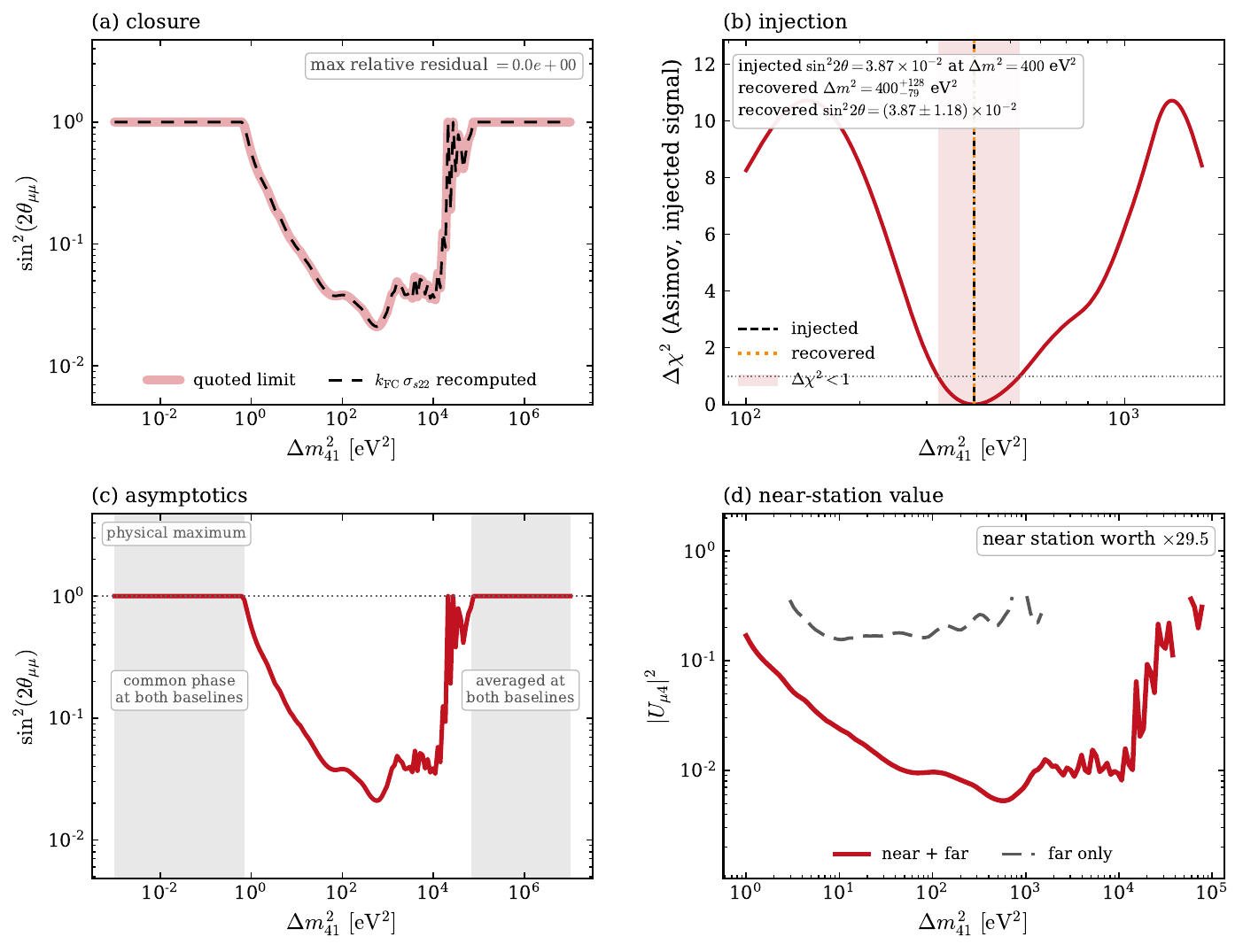}\hfill
  \includegraphics[width=0.49\textwidth]{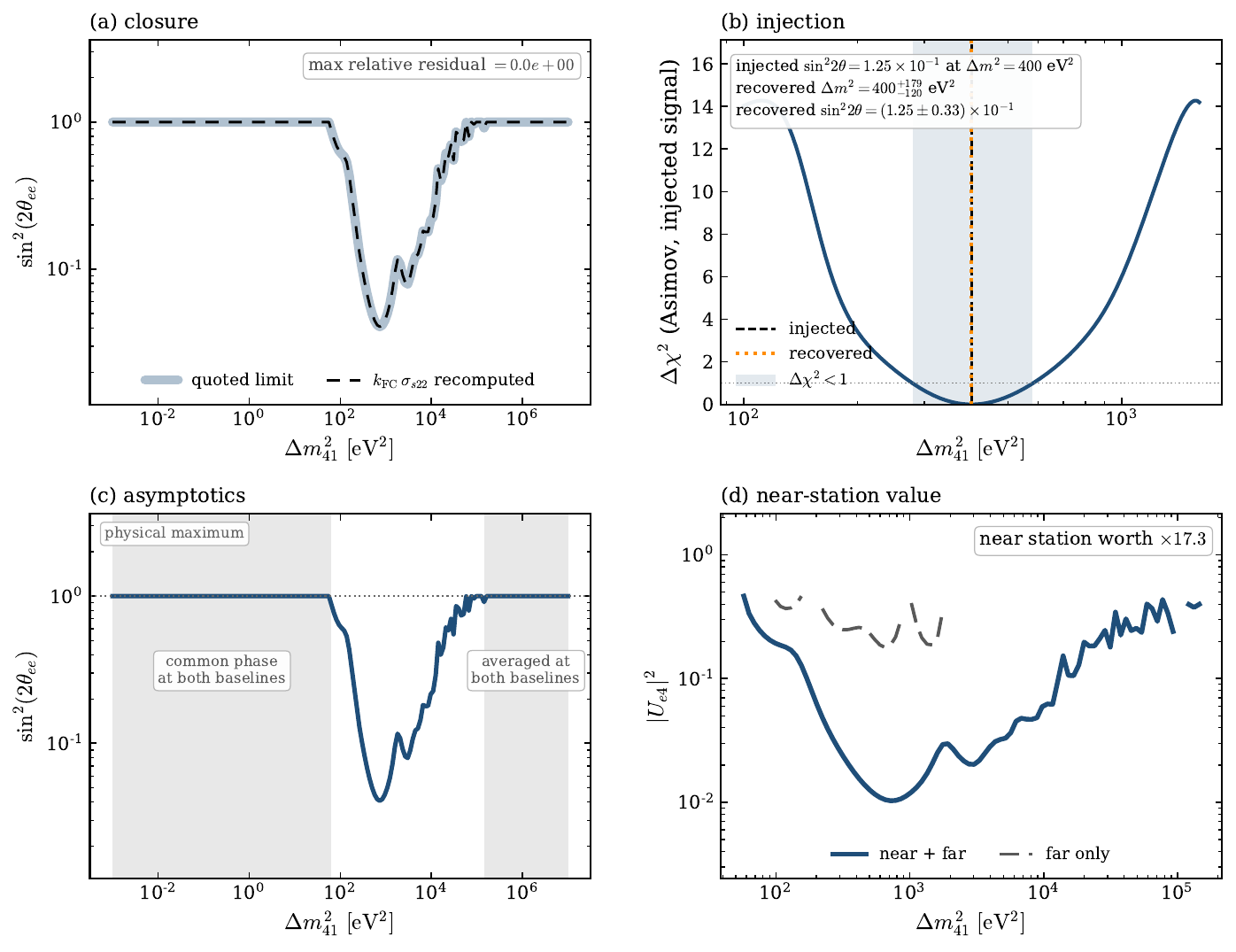}
  \caption{Validation of the muon (left) and electron (right) channels: (a)
  closure of the quoted limit against the recomputed belt, (b) recovery of an
  injected signal ($\Delta\chi^2$ profile in \dmsq{}), (c) the two asymptotic
  regimes, and (d) the sensitivity with and without the near station.
  \label{fig:valid}}
\end{figure*}

\section{Event samples}
\label{app:samples}

Figure~\ref{fig:vtx} shows the transverse distribution of the CC interaction
vertices in the two fiducial volumes: $68$, $90$, $95$ and $99\%$ of the
interactions lie within $17$, $20$, $22$ and \SI{25}{cm} of the axis at
the near station and within $59$, $72$, $78$ and \SI{87}{cm} at the far
station. Figure~\ref{fig:matched_all} compares the spectra at the two stations
with a common transverse acceptance and with the matched cone.

\section{Reconstruction details}
\label{app:reco}

\emph{Neutrino-energy response.} Figure~\ref{fig:nu_eres} summarises the
neutrino-energy response of the four samples as they enter the fit. The far
responses of both channels are strongly biased at the ends of the spectrum,
through containment for $\nue$ and through the multiple-scattering estimate of
the muon momentum for $\numu$; the migration matrices carry these biases in
full (Secs.~\ref{sec:nueE} and~\ref{sec:numufar}).

\emph{Muon momentum.} At the far station the muon momentum is estimated by a
linear least-squares fit of $\ln p_\mu$ to seven standardised features of the
muon track, whose points are digitised to the bar pitch of each layer: the
kink angle between straight-line fits to the first and last thirds of the
track, the r.m.s.\ residual of a straight-line fit to all points and to the
HCAL points alone, the depth reached in the HCAL, and the numbers of ECAL, HCAL
and total points. Multiple scattering in the
lead~\cite{Highland,PDG,Bolton:1997rm} makes the kink angle and the residuals
scale as $1/p$. Muons that do not reach the HCAL ($1.6\%$ of the sample) are
assigned \SI{0.5}{GeV}. The hadronic energy is a linear combination of the
visible ECAL and HCAL energies. Both estimators are calibrated on a random half
of the simulated events, and the far migration matrix is built from the
remaining, independent events. The quoted bias and core are the median and the
$68\%$ half-width of $E_{\mathrm{reco}}/E_{\mathrm{true}}-1$ in bins of true
energy. At the near
station the resolution follows the Gluckstern form~\cite{Gluckstern},
\begin{equation}
\left(\frac{\sigma_p}{p}\right)^2 =
\left(\frac{\sigma_x\,p}{0.3\,B\,L^2}\sqrt{\frac{720}{N+4}}\right)^2
+ \left(\frac{0.054}{B\sqrt{L\,X_0}}\right)^2,
\label{eq:gluckstern}
\end{equation}
with $\sigma_x$, $L$ and $X_0$ in metres, $B$ in tesla and $p$ in GeV$/c$; with
$B=\SI{1.9}{T}$, $\sigma_x=\SI{250}{\micro m}$ and $L=\SI{1.47}{m}$ of iron the
multiple-scattering term dominates at all momenta. The charge is determined
from the sign of the sagitta, correctly above about \SI{10}{GeV}.

\emph{Classifiers.} Table~\ref{tab:clf} summarises the classifiers. The
near-station variables are the gap between the vertex and the shower start, the
number of track hits at the vertex, the energy fraction and the ionisation (in
units of a minimum-ionising particle) at the shower start, the number of
electromagnetic clusters and the invariant mass of the two leading ones, the
transverse r.m.s.\ and the longitudinal length of the shower, its core energy
fraction, the electromagnetic and hadronic energies, the electromagnetic
fraction and the number of hits. All classifiers use trees of depth 4, a
learning rate of 0.05, a minimum child weight of 10 and $L_2$ regularisation
of 2, with the classes balanced by weight; the image classifiers in addition
subsample events (0.8) and features (0.3). The efficiencies used in the fit are
five-fold out-of-fold predictions. The near-station image classifier, on events
not used in its training, reaches $\OccSigEffOOF\%$ efficiency at
$\OccBkgRejOOF\%$ rejection and would not improve the $\nue$ limit. The classifier outputs are shown in
Fig.~\ref{fig:bdt_score}.

\begin{table}[!htb]
\centering
\caption{Classifiers for the $\nue$~CC versus NC-$\pizero$ separation.}
\label{tab:clf}
\begin{tabular}{lccc}
\toprule
 & Far & Near & Near, alt. \\
\midrule
Input & image & variables & image \\
Inputs & $45\,024$ & $13$ & $12\,003$ \\
$\nue$ CC events & $27\,578$ & $23\,199$ & $19\,880$ \\
NC-$\pizero$ events & $17\,800$ & $16\,462$ & $14\,178$ \\
\bottomrule
\end{tabular}
\end{table}

\section{Likelihood and validation}
\label{app:stat}

The near and far reconstructed spectra are stacked into one vector. For
$\nue$ the likelihood is evaluated on a common grid of 40 bins of
\SI{3.75}{GeV} per station, onto which each response matrix is mapped by
interval overlap, so that the energy information is that of the matrices' own
bins: \SI{7.5}{GeV} wide at the near station and four bins at the far station.
The prediction is $n_i(s,\theta)=n^0_i-s\,h_i+\sum_k\theta_kA_{ik}$, with
$s=\sstt_{\alpha\alpha}$, $n^0$ the null spectrum, $h$ its response to $s$ at
$s=0$, and $A$ the nuisance responses. In the Gaussian approximation on the
Asimov data set, with $W=\mathrm{diag}(1/n^0_i)$ and the nuisances profiled
analytically, the information on $s$ is
\begin{equation}
I = h^{\mathsf T} W h
- h^{\mathsf T} W A\,(A^{\mathsf T} W A + P^{-1})^{-1}A^{\mathsf T} W h,
\label{eq:fisher}
\end{equation}
where $P^{-1}=\mathrm{diag}(1/\sigma_k^2)$ is the prior precision (zero for a
free parameter) and $\sigma(s)=1/\sqrt{I}$. The median expected limit is
$s_{95}=\FCkMed\,\sigma(s)$, from the Feldman--Cousins belt of a Gaussian
estimator bounded at \mbox{$s\ge0$}, evaluated at each \dmsq{} separately with \dmsq{}
fixed (a raster scan); the $\pm1\sigma$ and $\pm2\sigma$ bands use factors
$\FCkLowOne$, $\FCkHighOne$ and $\FCkLowTwo$, $\FCkHighTwo$. The nuisance parameters are listed in
Table~\ref{tab:nuis}. With $2\times10^4$ pseudo-experiments per amplitude at five values of \dmsq{}
between $10^2$ and $10^4$~\si{eV^2}, Poisson-fluctuated and with the
constrained nuisances drawn from their priors, the factor $\FCkMed$ is
reproduced to within $2\%$ ($\numu$) and $4\%$ ($\nue$), never on the
optimistic side.

A signal injected at $\dmsq=\InjDmTrueMu$~\si{eV^2} with
$\sstt_{\mu\mu}=\InjAmpTrueMu$ is recovered at
$\InjDmLoMu$--$\InjDmHiMu$~\si{eV^2} with amplitude $\InjAmpMu\pm\InjAmpErrMu$;
in the $\nue$ channel an injection of $\sstt_{ee}=\InjAmpTrueE$ at the same
\dmsq{} is recovered at $\InjDmLoE$--$\InjDmHiE$~\si{eV^2} with amplitude
$\InjAmpE\pm\InjAmpErrE$. In Asimov data the recovery is exact by
construction; the intervals show the resolution in \dmsq{}
(Fig.~\ref{fig:valid}).

\begin{table}[!htb]
\centering
\caption{Nuisance parameters of the fits.}
\label{tab:nuis}
\begin{tabular}{llcc}
\toprule
Nuisance & Prior & $\numu$ & $\nue$ \\
\midrule
Common normalisation & free & 1 & 1 \\
Far/near normalisation & $1\%$ & 1 & 1 \\
Flux shape, Eq.~\eqref{eq:fluxcov} & unit/mode & 16 & 11 \\
NC-$\pizero$, per station & $30\%$ & -- & 2 \\
$\nutau\to e$, per station & $50\%$ & -- & 2 \\
\midrule
Total & & 18 & 17 \\
\bottomrule
\end{tabular}
\end{table}

\FloatBarrier
\bibliographystyle{apsrev4-2}
\bibliography{ship_nu}

\clearpage
\input{SHiP_authorlist_Sept26.tex}

\end{document}

%% file: numbers.tex
\newcommand{\CaloBkgRej}{\ensuremath{79.9}}
\newcommand{\CaloSigEff}{\ensuremath{83.2}}

\newcommand{\ConeRadiusWindowEnd}{\ensuremath{19.2}}
\newcommand{\DnineHigh}{\ensuremath{31.6}}
\newcommand{\DnineLow}{\ensuremath{12.7}}
\newcommand{\DnineNearOnlyHigh}{\ensuremath{81}}

\newcommand{\DnineShapeGapDeficitPct}{\ensuremath{9}}
\newcommand{\DnineShapeGapOne}{\ensuremath{17}}
\newcommand{\DnineShapeGapTwo}{\ensuremath{29}}

\newcommand{\DnineShapeSpanHigh}{\ensuremath{79.7}}
\newcommand{\DnineShapeSpanLow}{\ensuremath{13.7}}
\newcommand{\DnineWindowHigh}{\ensuremath{8.7}}
\newcommand{\DnineWindowLow}{\ensuremath{4.7}}

\newcommand{\FCkHighOne}{\ensuremath{2.96}}
\newcommand{\FCkHighTwo}{\ensuremath{3.96}}
\newcommand{\FCkLowOne}{\ensuremath{1.10}}
\newcommand{\FCkLowTwo}{\ensuremath{0.62}}
\newcommand{\FCkMed}{\ensuremath{1.96}}
\newcommand{\FarCompAl}{\ensuremath{7.9}}
\newcommand{\FarCompC}{\ensuremath{19.9}}
\newcommand{\FarCompH}{\ensuremath{1.4}}

\newcommand{\FarCompPb}{\ensuremath{70.8}}

\newcommand{\FarHadRes}{\ensuremath{36}}

\newcommand{\FarMuBiasFiftyHundred}{\ensuremath{-71}}

\newcommand{\FarMuBiasTwoFive}{\ensuremath{144}}
\newcommand{\FarMuCoreFiftyHundred}{\ensuremath{14}}
\newcommand{\FarMuCoreTwentyFifty}{\ensuremath{26}}

\newcommand{\FarMuShare}{\ensuremath{96}}

\newcommand{\FarNuMuECoreTwentyFifty}{\ensuremath{30}}

\newcommand{\FarNuMuERes}{\ensuremath{76}}

\newcommand{\FarOnlyMinMu}{\ensuremath{1.56\times10^{-1}}}

\newcommand{\FieldLossAll}{\ensuremath{7.6}}
\newcommand{\FieldLossFracFifty}{\ensuremath{3.2}}

\newcommand{\FieldLossFracTen}{\ensuremath{11.1}}

\newcommand{\FluxIndepPct}{\ensuremath{21}}
\newcommand{\InjAmpE}{\ensuremath{1.25\times10^{-1}}}
\newcommand{\InjAmpErrE}{\ensuremath{3.31\times10^{-2}}}
\newcommand{\InjAmpErrMu}{\ensuremath{1.18\times10^{-2}}}
\newcommand{\InjAmpMu}{\ensuremath{3.87\times10^{-2}}}
\newcommand{\InjAmpTrueE}{\ensuremath{1.25\times10^{-1}}}
\newcommand{\InjAmpTrueMu}{\ensuremath{3.87\times10^{-2}}}

\newcommand{\InjDmHiE}{\ensuremath{579}}
\newcommand{\InjDmHiMu}{\ensuremath{528}}
\newcommand{\InjDmLoE}{\ensuremath{280}}
\newcommand{\InjDmLoMu}{\ensuremath{321}}

\newcommand{\InjDmTrueMu}{\ensuremath{400}}
\newcommand{\LFar}{\ensuremath{97.09}}
\newcommand{\LNear}{\ensuremath{27.40}}

\newcommand{\LSpreadPct}{\ensuremath{9}}
\newcommand{\LSpreadPctE}{\ensuremath{4}}

\newcommand{\MCStatE}{\ensuremath{1.2}}

\newcommand{\MCStatMu}{\ensuremath{1.2}}

\newcommand{\NearChargeDevPct}{\ensuremath{3}}
\newcommand{\NearCompC}{\ensuremath{0.6}}

\newcommand{\NearCompSi}{\ensuremath{1.9}}
\newcommand{\NearCompW}{\ensuremath{97.5}}

\newcommand{\NearNueCalibN}{\ensuremath{3\,974}}

\newcommand{\NearNueNuCalCoreHigh}{\ensuremath{0.47}}
\newcommand{\NearNueNuCalCoreLow}{\ensuremath{0.32}}
\newcommand{\NearNueNuCalEmax}{\ensuremath{134}}
\newcommand{\NearNueNuCalEmin}{\ensuremath{13}}

\newcommand{\NearWindowPlates}{\ensuremath{71}}
\newcommand{\NearWindowTargetPct}{\ensuremath{59}}
\newcommand{\NearWindowVetoPct}{\ensuremath{0.6}}
\newcommand{\NearWorthE}{\ensuremath{17.3}}
\newcommand{\NearWorthMu}{\ensuremath{29.5}}

\newcommand{\NueNearBkgRej}{\ensuremath{81.7}}
\newcommand{\NueNearFrac}{\ensuremath{19.3}}
\newcommand{\NueNearSigEff}{\ensuremath{79.2}}
\newcommand{\NueNearZmax}{\ensuremath{27.82}}
\newcommand{\NueNearZmin}{\ensuremath{26.90}}
\newcommand{\OccBkgRejOOF}{\ensuremath{79.9}}
\newcommand{\OccSigEffOOF}{\ensuremath{73.5}}

\newcommand{\ProdZExpLambda}{\ensuremath{0.15}}

\newcommand{\RFN}{\ensuremath{0.101}}
\newcommand{\RFNE}{\ensuremath{0.100}}

\newcommand{\SDIronSampling}{\ensuremath{5.0}}

\newcommand{\SDPlanesPerSection}{\ensuremath{14}}

\newcommand{\SNDFeZmaxOne}{\ensuremath{29.225}}
\newcommand{\SNDFeZmaxThree}{\ensuremath{30.933}}
\newcommand{\SNDFeZmaxTwo}{\ensuremath{30.079}}
\newcommand{\SNDFeZminOne}{\ensuremath{28.382}}
\newcommand{\SNDFeZminThree}{\ensuremath{30.090}}
\newcommand{\SNDFeZminTwo}{\ensuremath{29.236}}
\newcommand{\SNDFibreDiam}{\ensuremath{250}}
\newcommand{\SNDFibreLayerThick}{\ensuremath{0.5}}

\newcommand{\SNDPlateThick}{\ensuremath{3.5}}

\newcommand{\SNDTargetPlates}{\ensuremath{120}}

\newcommand{\SNDTargetZmax}{\ensuremath{28.372}}
\newcommand{\SNDTargetZmin}{\ensuremath{27.016}}
\newcommand{\SNDTileSize}{\ensuremath{5}}
\newcommand{\SNDTileThick}{\ensuremath{10}}
\newcommand{\SNDVetoZmax}{\ensuremath{26.996}}
\newcommand{\SNDVetoZmin}{\ensuremath{26.956}}

\newcommand{\ScaleCompHalfMu}{\ensuremath{1.60}}

\newcommand{\ScaleCompOneMu}{\ensuremath{1.80}}

\newcommand{\ShapeOnePctE}{\ensuremath{1.1}}
\newcommand{\ShapeOnePctMu}{\ensuremath{2.4}}
\newcommand{\SigRelFiveE}{\ensuremath{2.83}}
\newcommand{\SigRelFiveMu}{\ensuremath{1.02}}
\newcommand{\SigRelTwoE}{\ensuremath{1.55}}
\newcommand{\SigRelTwoMu}{\ensuremath{1.02}}

\newcommand{\SpecFarResCalo}{\ensuremath{76}}

\newcommand{\SpecFarResSpec}{\ensuremath{30}}

\newcommand{\SpecResolution}{\ensuremath{1}}

\newcommand{\SpecScaleCostMu}{\ensuremath{1.23}}

\newcommand{\ThetaMax}{\ensuremath{6.886}}

\newcommand{\UeLimit}{\ensuremath{1.03\times10^{-2}}}
\newcommand{\UeLimitDm}{\ensuremath{716}}

\newcommand{\UeLimitMass}{\ensuremath{26.8}}

\newcommand{\UeLimitRelTwo}{\ensuremath{1.60\times10^{-2}}}

\newcommand{\UmuLimit}{\ensuremath{5.28\times10^{-3}}}

\newcommand{\UmuLimitDm}{\ensuremath{598}}

\newcommand{\UmuLimitMass}{\ensuremath{24.4}}
\newcommand{\UmuLimitNoShape}{\ensuremath{1.35\times10^{-3}}}

\newcommand{\UmuLimitRelTwo}{\ensuremath{5.38\times10^{-3}}}

\newcommand{\UmuLimitSpecFine}{\ensuremath{3.02\times10^{-3}}}

\newcommand{\UmuLimitSpecNu}{\ensuremath{3.08\times10^{-3}}}

\newcommand{\UmuLimitSpecNubar}{\ensuremath{4.07\times10^{-3}}}

%% file: table1.tex
\begin{table*}[t]
\centering
\caption{Expected yields at $6\times10^{20}$~PoT in the matched angular acceptance and the fitted energy range, before and after selection, as they enter the fit. Near: the silicon-tungsten target-vertex fiducial (26.90--27.82~m) for both samples; far: the calorimeter window. Selected $\nu_e$ CC and $\nu_\mu$ CC are the signals of the two samples; $\nu_\tau\to e$ and NC-$\pi^0$ are the selected backgrounds of the $\nu_e$ sample. 1.4--2.5\% of the CC interactions lie outside the fitted energy range and are not included.}
\label{tab:yields}
\begin{tabular}{llcccc}
\toprule
 & & \multicolumn{2}{c}{Near} & \multicolumn{2}{c}{Far} \\
\cmidrule(lr){3-4}\cmidrule(lr){5-6}
Interactions & Sample & In acceptance & Selected & In acceptance & Selected \\
\midrule
$\nue+\bar\nu_e$ CC & $\nu_e$ & $8.18\times10^{5}$ & $6.42\times10^{5}$ & $8.19\times10^{4}$ & $7.39\times10^{4}$ \\
$\numu+\bar\nu_\mu$ CC & $\nu_\mu$ & $1.85\times10^{6}$ & $1.67\times10^{6}$ & $1.87\times10^{5}$ & $1.68\times10^{5}$ \\
$\nutau+\bar\nu_\tau$ CC, $\tau\to e$ & $\nu_e$ (bkg.) & $6.86\times10^{3}$ & $5.24\times10^{3}$ & $6.94\times10^{2}$ & $6.24\times10^{2}$ \\
NC-$\pizero$ & $\nu_e$ (bkg.) & $5.59\times10^{5}$ & $9.86\times10^{4}$ & $5.29\times10^{4}$ & $1.04\times10^{4}$ \\
\bottomrule
\end{tabular}
\end{table*}

%% file: SHiP_authorlist_Sept26.tex


\onecolumngrid
\noindent

\centerline{\Large\bf The SHiP Collaboration}
\vspace*{1mm}
\noindent
\begin{center}
M.~Akincy$^{39}$\orcidlink{0000-0003-4879-8827},
A.~Aksoy$^{13}$\orcidlink{0009-0001-3734-2157},
R.~Albanese$^{20,d}$\orcidlink{0000-0003-4586-8068},
K.~Albrecht$^{9}$\orcidlink{0000-0001-5257-3474},
F.~Alessio$^{35}$\orcidlink{0000-0001-5317-1098},
A.~Alexandrov$^{20,d}$\orcidlink{0000-0002-1813-1485},
F.~Aloschi$^{20,d}$\orcidlink{0000-0002-2501-7525},
A.~Anokhina$^{27}$\orcidlink{0000-0002-4654-4535},
T.~Arndt$^{14}$\orcidlink{0000-0002-7797-8862},
D.~Arutinov$^{13}$\orcidlink{0000-0001-6150-7030}
% ...
\end{center}
\vspace*{0.5cm}

{\footnotesize \it

\noindent
% ...
$^{9}$Institut f\"{u}r Physik, Humboldt-Universit\"{a}t zu Berlin, Berlin, Germany\\
% ...
$^{13}$Integrated Computing Architectures (ICA | PGI-4),Forschungszentrum J\"{u}lich GmbH (KFA),  J\"{u}lich , Germany\\
$^{14}$Karlsruhe Institute of Technology, Karlsruhe, Germany \\
% ...
$^{20}$Sezione INFN di Napoli, Napoli, Italy\\
% ...
$^{27}$Institute of Nuclear Physics, Almaty, 
Kazakhstan\\
% ...
$^{35}$European Organization for Nuclear Research (CERN), Geneva, Switzerland\\
% ...
$^{39}$Ankara University, Ankara, Turkey\\
% ...
$^{d}$Universit\`{a} di Napoli ``Federico II``, Napoli, Italy \\
% ...
}